\documentclass[9pt]{article}  \usepackage{times}
\usepackage{graphicx, textgreek}

\usepackage{color}

\usepackage[round,numbers,sort&compress]{natbib} 

\usepackage{amsmath}
\usepackage{nccmath}   
\usepackage{amssymb}
\usepackage{stackrel}
\usepackage{chemarrow}
\usepackage{graphicx}
\usepackage{textgreek}

\usepackage{booktabs}
\usepackage{dcolumn}
\usepackage{rotating}
\usepackage{multirow}

\renewcommand\appendix{\par
	\setcounter{section}{0}
	\setcounter{subsection}{0}
	\setcounter{figure}{0}
	\setcounter{table}{0}
	\renewcommand\thesection{Appendix \Alph{section}}
	\renewcommand\thefigure{\Alph{section}\arabic{figure}}
	\renewcommand\thetable{\Alph{section}\arabic{table}}
}

\begin{document}

	\twocolumn[{\LARGE \textbf{The mechanical properties of nerves, the size of the action potential, and consequences for the brain\\*[0.2cm]}}
	{\large Thomas Heimburg$^\ast$\\*[0.1cm]
		{\small Niels Bohr Institute, University of Copenhagen, Jagtvej 128, 2200 Copenhagen N, Denmark}\\*[-0.1cm]
		
		{\normalsize \textbf{ABSTRACT}\hspace{0.5cm} The action potential is widely considered a purely electrical phenomenon. However, one also finds mechanical and thermal changes that can be observed experimentally. In particular,  nerve membranes become thicker and axons contract. The spatial length of the action potential can be quite large, ranging from millimeters to many centimeters. This suggests to employ macroscopic thermodynamics methods to understand its properties. The pulse length is several orders of magnitude larger than the synaptic gap, larger than the distance of the nodes of Ranvier, and even larger than the size of many neurons such as pyramidal cells or brain stem motor neurons. Here, we review the mechanical changes in nerves, theoretical possibilities to explain them, and implications of a mechanical nerve pulse for the neuron and for the brain. In particular, the contraction of nerves gives rise to the possibility of fast mechanical synapses.
			\\*[0.3cm] }}
	\noindent\footnotesize{\textbf{Keywords:} Nerves ; pulse velocity ; pulse length ; mechanical changes ; nerve contraction ; solitons 
\\*[0.1cm]}
	\noindent\footnotesize {$^{\ast}$corresponding author, theimbu@nbi.ku.dk. }\\
	\vspace{0.3cm}
	]

	\normalsize


\section{Introduction}
\label{introduction}

In neurophysiology, the nerve pulse is considered a purely electrical phenomenon. Since the electrical signals in nerve axons display an amplitude of about 100 mV and a duration of a few milliseconds, it is easy to measure the action potentials even with simple electronics. However, there exist other variables that also change during the action potential such as the temperature, the thickness and the length of nerves. These changes are much more difficult to measure because they are of the order of micro-Kelvin or nanometers, respectively, and therefore seemingly small. For this reason, these changes are often considered marginal side-effects that can be neglected. However, the ease of measurement does not permit a judgment about the magnitude of a phenomenon if one compares measurements that carry different units. While the temperature change during the action potential seems small and the voltage change seems large, this might not be the case for the associated energies. In fact, it has been shown that the thermal energy of an action potential at the pulse maximum, which is given by $\Delta Q=c_p \Delta T$, is larger than the electrical energy given by $\Delta E=\frac{1}{2}C_m (V-V_0)^2$, where $V_0$ is the resting potential \cite{Howarth1968, Ritchie1985}. The changes in heat and voltage are proportional and in phase, see Fig. \ref{heat_voltage_ritchie1985}. Therefore, it is likely that they represent aspects of the same phenomenon. In the past, this lead to much confusion about the true nature of the nervous impulse (reviewed in \cite{Heimburg2021}).

The heat of a nerve pulse is indeed an interesting and poorly understood phenomenon. In the 1840s, Hermann von Helmholtz studied various properties of frog muscles and nerves. He found that during the contraction of muscles heat is generated, while during the propagation of the nerve pulse no such heat exchange can be found \cite{Helmholtz1848}. Helmholtz found this surprising because the ion composition of muscles and nerves from frogs is quite similar. The absence of any heat exchange bothered a complete generation of neuroscientists in the second half of the 19$^{th}$ century. Their efforts were reviewed in a paper by the later Nobel prize laureate Archibald V. Hill \cite{Hill1912}. Hill summarized the state of the art in 1912 as follows: \emph{``This suggests very strongly, {\ldots} , that the propagated nervous impulse is not a wave of irreversible chemical breakdown, but a reversible change of a purely physical nature''}. Bayliss \cite{Bayliss1915} stated in 1915: \emph{``The result makes it impossible to suppose that any chemical process resulting in an irreversible loss of energy can be involved in the transmission of a nerve impulse, and indicates that a reversible physicochemical one of some kind is to be looked for.''} There have been various more recent reports about the heat production in nerve \cite{Abbott1958, Abbott1965, Howarth1968, Abbott1973, Howarth1975, Ritchie1985, Tasaki1981, Tasaki1989}, which were thoroughly reviewed in \cite{Heimburg2021}. A higher sensitivity and time resolution of the instrumentation allowed to measure the heat production during the nerve pulse more accurately. It is now known that during the depolarization phase of the action potential heat is released into the environment of the nerve, which is reabsorbed in the repolarization phase \cite{Abbott1958}. The integral over the total heat exchange rate is zero within experimental accuracy. No heat is dissipated after the passage of the nerve pulse \cite{Ritchie1985}. The reversible transfer of heat into the environment is unexpected in traditional electrical nerve models that rely on irreversible processes, as already noted by A.L. Hodgkin in his monograph ``The conduction of the nervous impulse'' \cite{Hodgkin1964}. A thermodynamic consideration of heat changes in nerves in a purely electrical description would predict observing temperature changes that correspond to the temporal change of the electrical energy on the membrane capacitor (e.g., \cite{Howarth1968, Howarth1975}). However, the magnitude of the capacitive energy is much smaller than the heat exchange that is observed. For this reason, a purely electrical model cannot explain the heat production in nerves. Howarth et al. \cite{Howarth1968} proposed that the heat production most likely is due to reversible changes in entropy, corresponding for instance to changes in the molecular structure of the membrane itself. Wei propose in 1972 that the reversible heat release may be related to changes in the dipole orientation of membrane components. Thus, the reversible heat changes are consistent with a reversible transition in the physical state of the membrane or membrane molecules. The above consideration lead to two conclusions:

\begin{figure}[htbp]
\centering
\includegraphics[width=225pt,height=172pt]{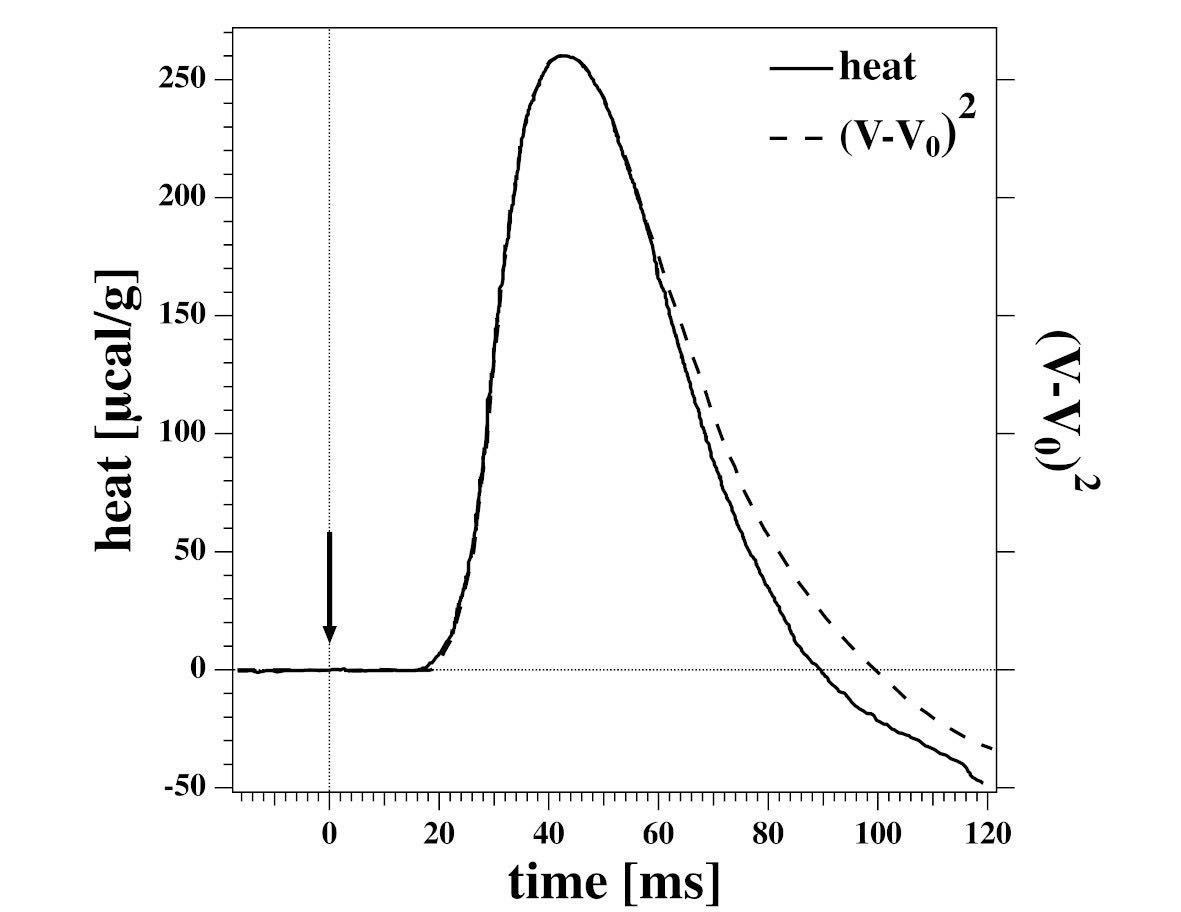}
\caption{The heat exchanged with the environment in garfish olfactory nerve nerve compared to the electrical energy which is proportional to $(V-V_0)^2$, where $V_0$ is the resting potential. The time courses of the heat and the electrical energy are very similar, but they display different magnitude. Adapted from \cite{Ritchie1985}.}
\label{heat_voltage_ritchie1985}
\end{figure}

\begin{enumerate}
\item The nerve pulse contains non-electrical contributions because the total amount of heat released and reabsorbed is larger than the electrical energy.

\item Since no net heat is exchanged after the completion of the pulse, the nerve pulse is isentropic.

\end{enumerate}

A typical example for an isentropic process is a mechanical wave. This is the reason why Hill \cite{Hill1912} proposed that the nerve pulses is a reversible physical phenomenon such as the propagation of sound. The sound velocity in nerve membranes is of the order of 170--180 m\slash s if only the motion of the membrane itself is considered in a calculation \cite{Heimburg2005c}, while it can be as low as 1 m\slash s if the surface-associated water is considered \cite{Kappler2017, Schneider2021}. This is exactly the range of the propagation velocity in nerves that can be of the order of 100 m\slash s for myelinated motor neurons (e.g., 70 m\slash s for the sensory compound action potential in the median nerve in humans \cite{Krarup1994}) and as low as 1 m\slash s in the femoral nerve of locust \cite{Villagran2011} and density waves in lipid monolayers \cite{Griesbauer2012a, Griesbauer2012b}.

\begin{figure}[htbp]
\centering
\includegraphics[width=225pt,height=117pt]{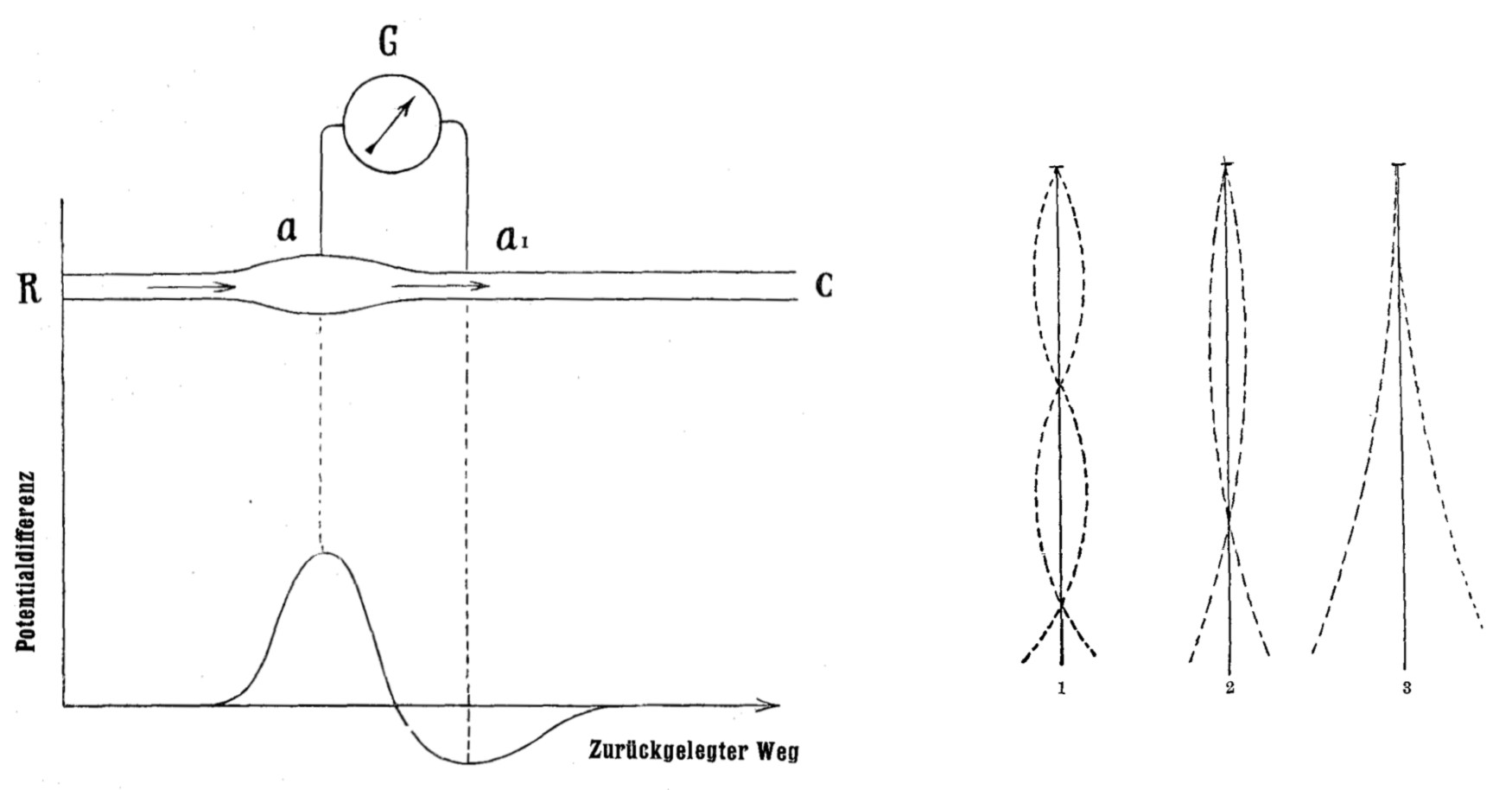}
\caption{The experiment by Wilke and Atzler \cite{Wilke1912a, Wilke1912b}. Left: Schematic drawing of the propagation of a peristaltic pulse in a nerve compared to the distance traveled. Right: The motion of thin glass fibers attached to the end of the nerve indicate a contraction of the excited nerve under the influence of an action potential. Figure from \cite{Wilke1912a} (left) and \cite{Wilke1912b} (right).}
\label{wilke1912}
\end{figure}

The above discussion suggests that the nerve pulse has mechanical aspects. Already in 1880 it was well-known that nerves can be excited mechanically \cite{Tigerstedt1880}. To our knowledge, the first direct experiments demonstrating mechanical aspects during nerve pulse propagation were performed by Wilke and Atzler \cite{Wilke1912a, Wilke1912b} who studied the contraction of frog nerves. The starting hypothesis of their experiment was the possibility of deformation waves in the nerve axons. Considering a peristaltic excitation traveling along a nerve cylinder containing an incompressible fluid, they expected the cylinder to contract. Since Wilke and Atzler expected that thickness changes in the axons were very small, they rather focused on the contraction of the nerve because it integrates the mechanical changes over the whole length of the axon. They attached a very thin glass fiber to the end of a nerve and found it to vibrate after the stimulation of the nerve (see Fig. \ref{wilke1912}). They concluded that the nerve pulse is associated with mechanical changes, and that nerve contraction can indeed be demonstrated.

Mechanical changes during the nerve pulse may in fact be of physiological relevance. Many recent investigations studied the excitation of the brain by focussed ultrasound \cite{Tufail2011, Tyler2012, Mueller2014, Rezayat2016, Ye2016, Kamimura2016, Tyler2018, Blackmore2019, Kamimura2020, Feng2024}. It is hard to understand why ultrasound can excite nerves if the underlying mechanism was purely electrical. This largely un-understood phenomenon may find an explanation if a nerve pulse had a mechanical component. Recent magnetic resonance experiments indicate that active brain regions increase their stiffness \cite{Palnitkar2024}, which suggest changes in the elastic constants. Mechanical activity of stimulated neurons also may also help understanding the increase of the stimulation threshold of nerves upon stretching \cite{Heimburg2022a}.

In this article we review some of the evidence for mechanical changes in nerves from todays point of view, and consequences for the function of nerves and the nervous system.


\section{Velocity and length of the nerve pulse}
\label{velocityandlengthofthenervepulse}

\begin{figure}[htbp]
\centering
\includegraphics[width=225pt,height=77pt]{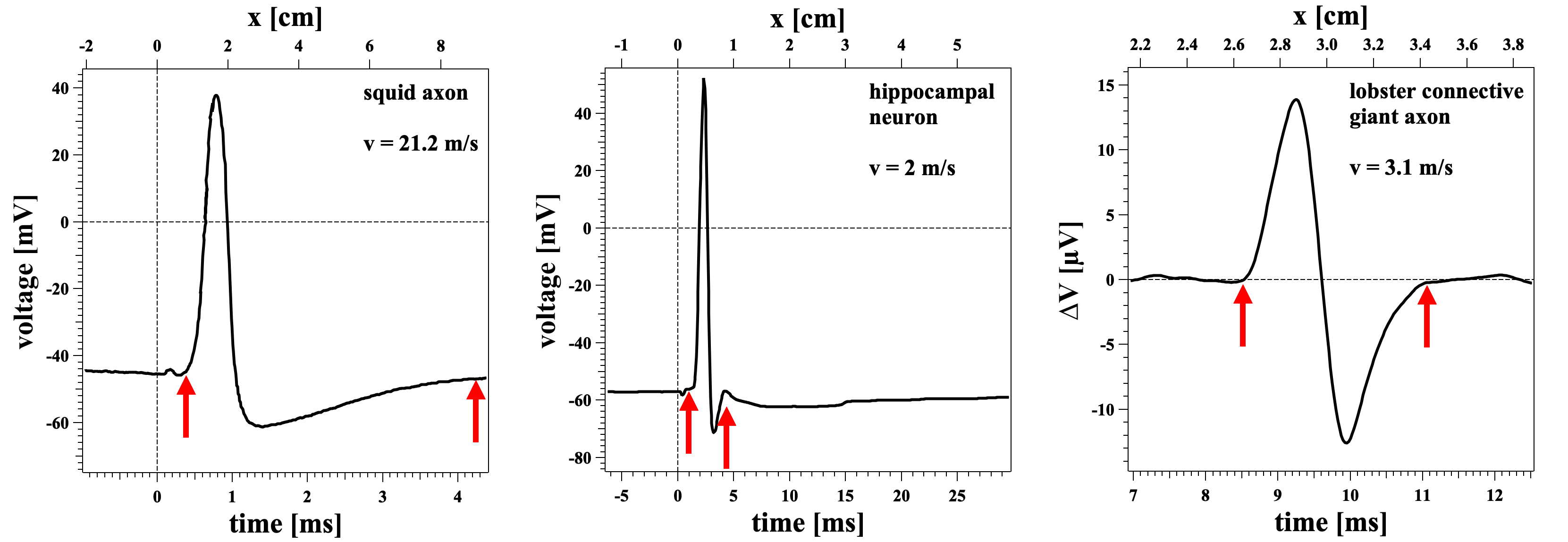}
\caption{Time- and length-scale of action potentials in single neurons. The length of the pulse is determined between the arrows. Left: Action potential of a squid axon at 20$^\circ$C. The velocity is 21.2 m\slash s and the length of the pulse is $\Delta x=8.2$ cm. Adapted from \cite{Hodgkin1939}. Center: A fast-firing medial-septum-diagonal band hippocampal neurons from rat brain. The velocity of the pulse is 2 m\slash s and the length of the pulse is $\Delta x=0.7$ cm. Adapted from \cite{Jones1999}. Right: Action potential of a giant axon from the lobster connectives. The velocity is 3.1 m\slash s and the length of the pulse is $\Delta x=0.8$ cm. Adapted from \cite{GonzalezPerez2016}.}
\label{velocity_single_neurons}
\end{figure}

\begin{figure}[htbp]
\centering
\includegraphics[width=225pt,height=76pt]{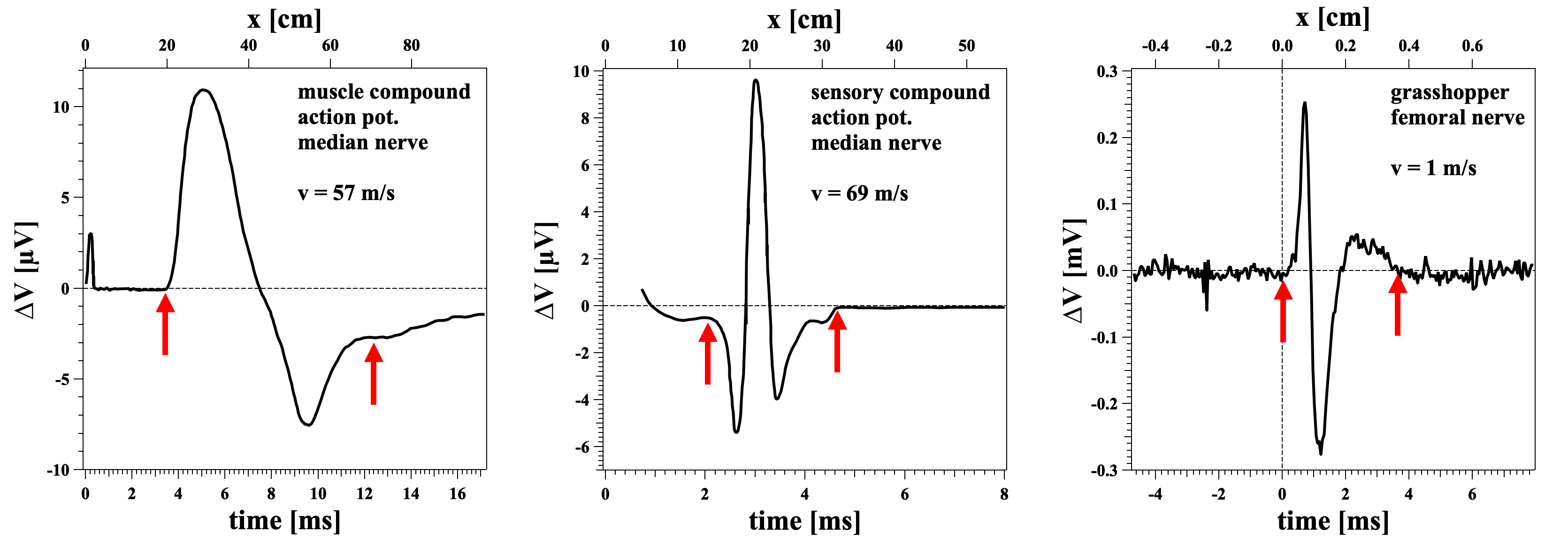}
\caption{Time- and length-scale of compound action potentials in nerves. Left: Compound muscle action potential from the median nerve in humans. The velocity is 57 m\slash s and the length of the pulse is $\Delta x=51$ cm. Adapted from \cite{Krarup1994}. Center: Compound sensory action potential (CSAP) of the median nerve. The velocity is 69 m\slash s and the length of the pulse is $\Delta x=18$ cm. Adapted from \cite{Krarup1994}. Right: Compound action potential of the femoral nerve in grasshopper (locust) legs. The velocity is 1m\slash s and the length of the pulse is $\Delta x=0.37$ cm. Adapted from \cite{Villagran2011}. }
\label{velocity_nerves}
\end{figure}

First, we show that the nerve pulse is macroscopic and that for this reason macroscopic thermodynamic treatment of the nerve membrane properties is justified. We have already mentioned that typical velocities of nerves pulses are in the range of 1--100 m\slash s. The physical length $\Delta x$ of the action potential in space is given by
\begin{equation}\label{eq:length01}
	\Delta x = v \cdot \Delta t \;,
\end{equation}
where $v$ is the pulse velocity and $\Delta t$ is the duration of the nerve pulse. For a motor neuron with a pulse velocity of 100 m\slash s and a duration of 4 ms, this yields 40 cm, i.e., a significant part of the overall length of the motor neuron itself (which can be one meter long in humans). In Figs. \ref{velocity_single_neurons} and \ref{velocity_nerves} we show the physical dimension of action potentials in several neurons and nerves as calculated by eq. \ref{eq:length01}. The beginning and the end of the nerve pulse in time are given by the red arrows. The results are listed in table \ref{tab:prop_nerv_1}. One finds that the length of the action potential varies between 4 mm (locust femoral nerve) to 51 cm in the human median nerve. Most action potentials are similar in duration but they can display total lengths that are quite different due to quite different propagation velocities. A typical ion channel protein has a diameter of about 5 nm. The length of the actual nerve pulse is thus up to eight orders of magnitude larger than that of a single channel protein, which is the central element in electrical models for the nerve pulse. It is also several orders of magnitude larger than the typical size of a synaptic gap of a chemical synapse (20--40 nm \cite{Hormuzdi2004}). In myelinated motor neurons it is much larger than the distance between nodes of Ranvier (about one millimeter, \cite{Kandel2000}) which raises questions concerning the validity of the concept of saltatory conduction.

\begin{table}[htb!]
\def\arraystretch{0.5}
\begin{tabular*}{8.5cm}{lcccc}
\hline
\scriptsize system & \scriptsize velocity [m/s] & \scriptsize $\Delta t$ [ms] & \scriptsize $\Delta x$ [cm] & \scriptsize source\\
\hline
\scriptsize {\textbf single neurons: }& & & & \\
\scriptsize squid axon &\scriptsize 21.2 & \scriptsize 3.9 & \scriptsize 8.2 & \scriptsize \cite{Hodgkin1939}\\
\scriptsize hippocampal fast-firing neuron&\scriptsize 2 & \scriptsize 3.5 & \scriptsize 0.7 & \scriptsize \cite{Jones1999}\\
\scriptsize giant axon lobster connective &\scriptsize 3.1 & \scriptsize 2.6 & \scriptsize 0.81 & \scriptsize \cite{Jones1999}\\
\scriptsize {\textbf nerves:} & &   &   &  \\
\scriptsize median nerve  CMAP &\scriptsize 57 & \scriptsize 9.0 & \scriptsize 51.4 & \scriptsize \cite{Krarup1994}\\
\scriptsize median nerve  CSAP &\scriptsize 69 & \scriptsize 2.6 &  \scriptsize 17.9 & \scriptsize \cite{Krarup1994}\\
\scriptsize locust femoral &\scriptsize 1 & \scriptsize 3.65 &  \scriptsize 0.37 & \scriptsize \cite{Villagran2011}\\
\scriptsize pike olfactory &\scriptsize 0.042 & \scriptsize 270 &\scriptsize 1.13& \scriptsize \cite{vonMuralt1976}\\
\hline
\end{tabular*}
\caption{\small Velocity, timescale and spatial extension of action potentials in neurons and nerves determined from Figs. \ref{velocity_single_neurons} and \ref{velocity_nerves}, and the extremely slow action potential from pike olfactory nerve at 0$^\circ$ C. CMAP denotes the compound motor action potential, and CSAP the compound sensory action potential.}
\label{tab:prop_nerv_1}
\end{table}

\begin{figure}[htbp]
\centering
\includegraphics[width=225pt,height=87pt]{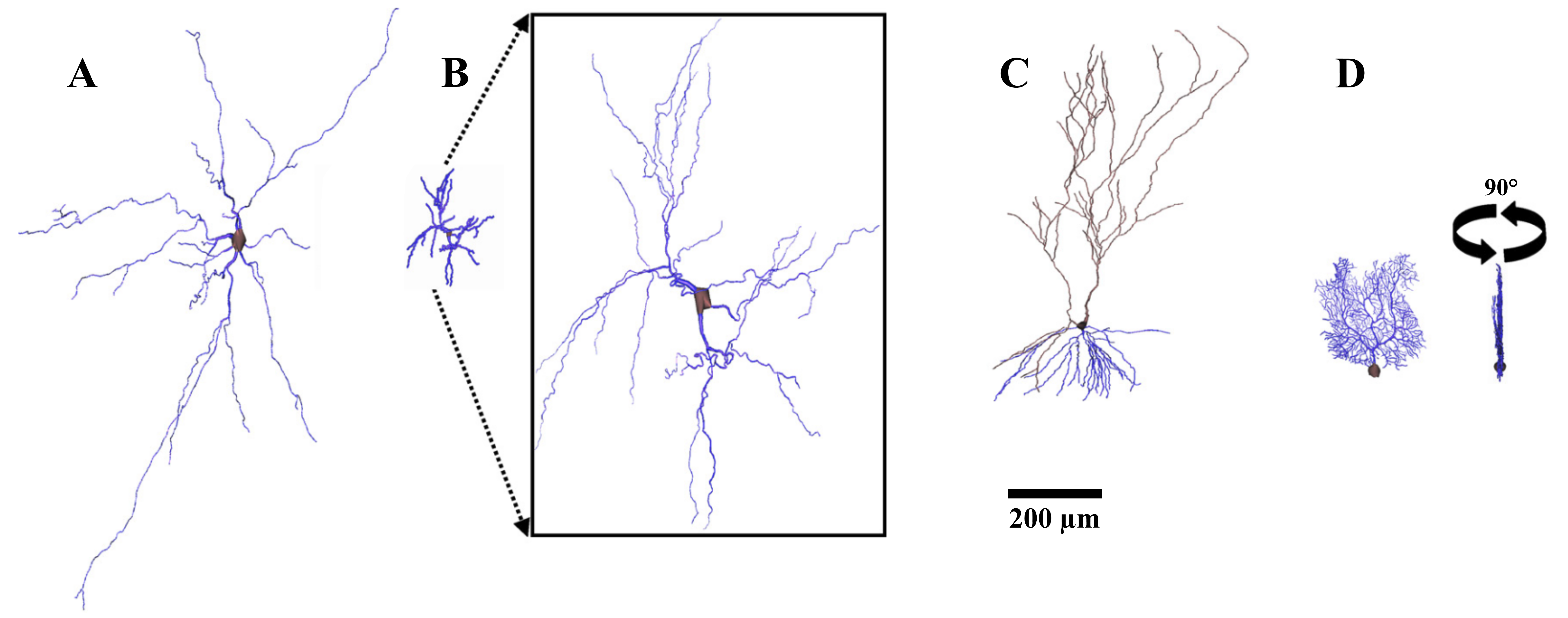}
\caption{Size and shape of neurons from rat brains. A: Brainstem motoneuron with diameter of 1.4 mm. B: Stellate cell with a diameter of 230 \textmu m. The insert box is the same cell amplified 4 times. C: Hippocampal CA3 pyramidal cell with a diameter of 850 \textmu m. D: Purkinie cell with a diameter of 230 \textmu m. The scale bar is 200 \textmu m. Adapted from \cite{Brown2008} with permission.}
\label{brown_et_al_2008_nerves_rearranged}
\end{figure}

\begin{figure}[htbp]
\centering
\includegraphics[width=169pt,height=169pt]{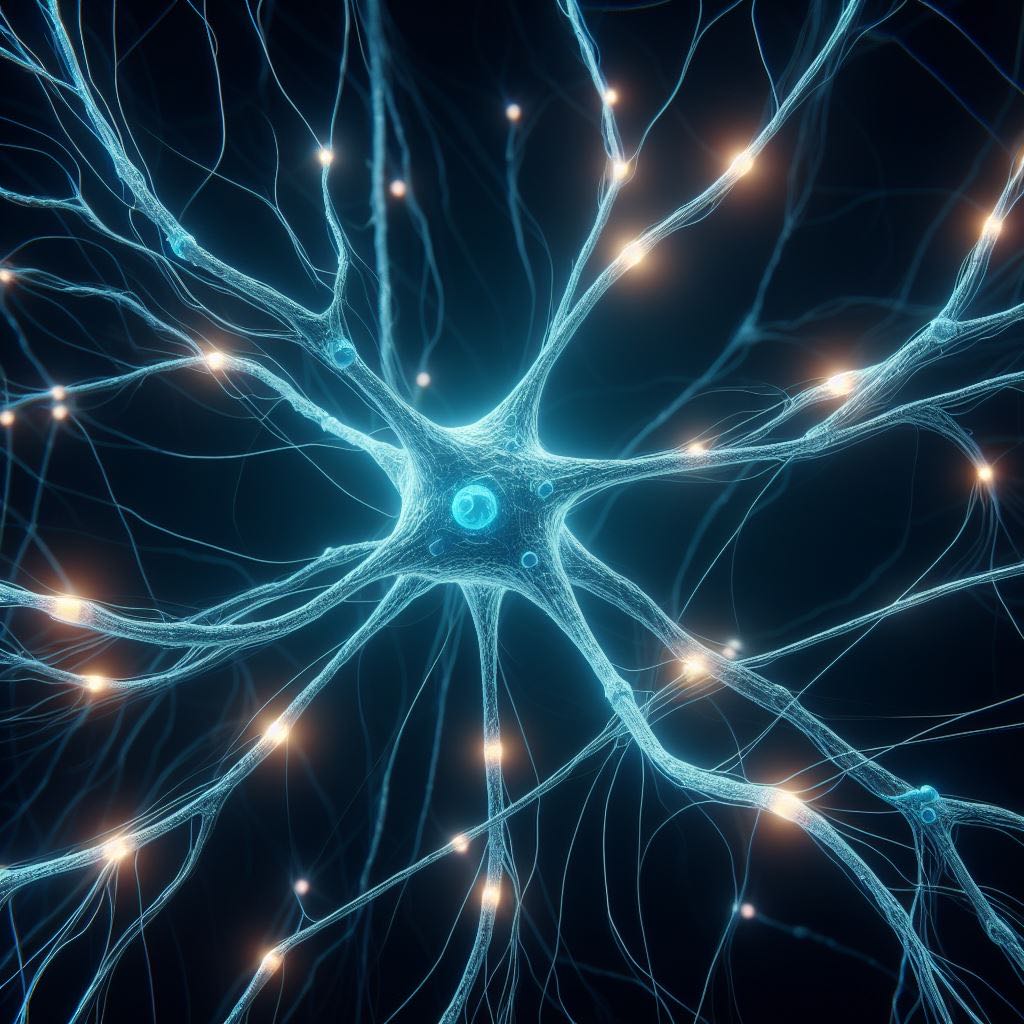}
\caption{The common artistic view of nerve pulses (bright spots on the dendrites) in neural networks is misleading. It represents action potentials as a microscopic events of similar dimensions as the diameter of dendrites. Similar pictures of neurons in the brain are often used by neuroscience groups in internet presentations and in popular publications. However, action potentials have dimensions that are orders of magnitude larger than the bright regions in this image (see text). The image was AI-generated with Microsoft Image Creator based on DALL-E3.}
\label{size_neuron_bing}
\end{figure}

The nerve pulse may also be much larger than many single neurons. Fig. \ref{brown_et_al_2008_nerves_rearranged} shows four different neurons: a brainstem motor neuron (maximal length 1.4 mm), a stellate cell (230 \textmu m), a hippocampal pyramidal cell (850 \textmu m) and a Purkinie cell (230 \textmu m) - all of them significantly smaller than a typical single action potential (cell images from \cite{Brown2008}). This implies that an action potential can span over several neurons or even regions of neurons. Thus, artistic representations of nerve pulses in neural nets as shown in Fig. \ref{size_neuron_bing} (such pictures can be found in many popular articles and on the websites of neuroscience groups) are quite misleading because they suggests that the size of a nerve pulse is similar to the diameter of dendrites (i.e., \textmu m scale), which is clearly incorrect. Such a picture overemphasizes the importance of cellular detail over macroscopic properties. The appreciation of the physical length of pulses underlines the need for a macroscopic thermodynamic description. This will rather rely on macroscopic susceptibilities such as compressibility and heat capacity, intensive variables such as pressure, temperature, voltage and the chemical potentials of membrane components, and on extensive properties such as membrane charge, the entropy of the nerve membrane, and the length and thickness of the axonal membranes.

\section{Experimental evidence for mechanical changes in nerves}
\label{experimentalevidenceformechanicalchangesinnerves}


\subsection{Mechanical recordings of the action potential: Thickness changes}
\label{mechanicalrecordingsoftheactionpotential:thicknesschanges}

\begin{figure}[htbp]
\centering
\includegraphics[width=225pt,height=91pt]{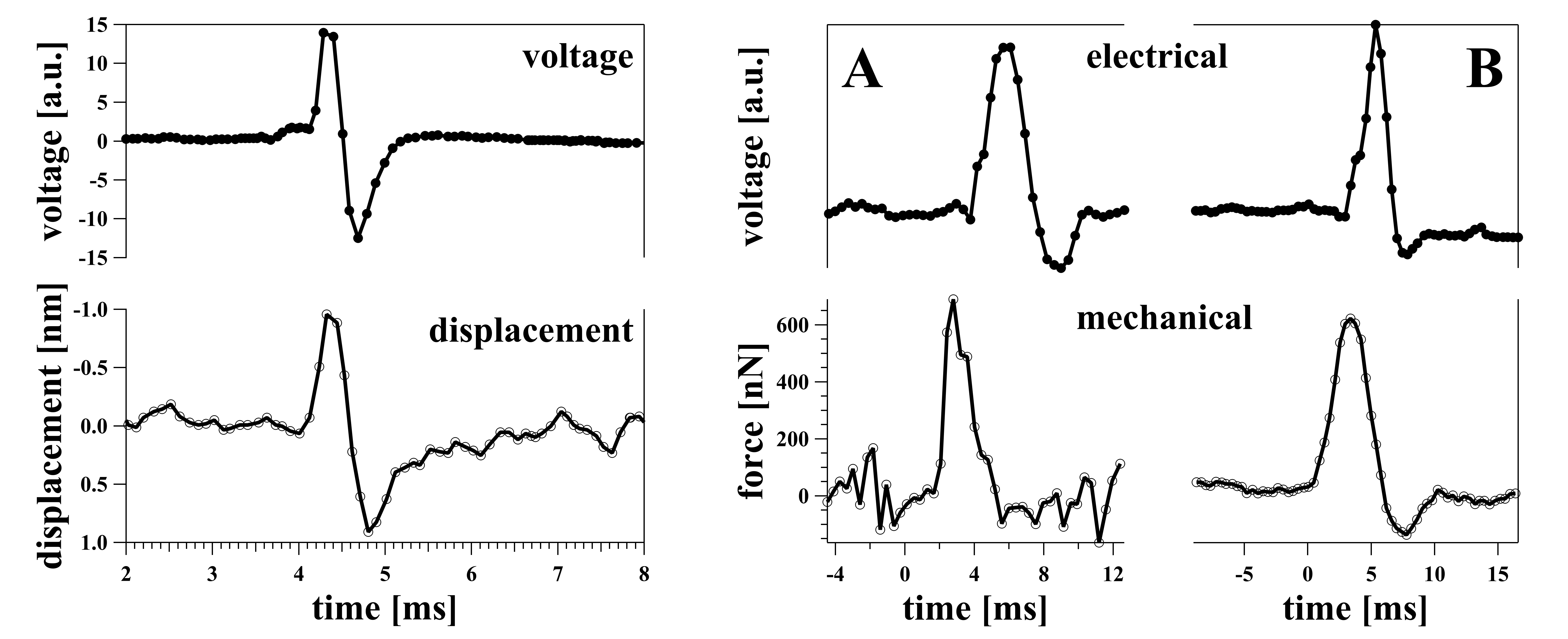}
\caption{Left: Thickness change (vertical dislocation) of a squid axon (bottom) under the influence of the action potential (top) as measured by a mechanical cantilever. The magnitude of the thickness change is about 1nm. Adapted from \cite{Iwasa1980a}. Right: The force of crab nerve fibers excreted on a cantilever. Two different experiments are shown Adapted from \cite{Iwasa1980b}.}
\label{iwasa_tasaki_a_b}
\end{figure}

Mechanical changes in nerve thickness and length can be measured by atomic force microscope cantilevers, or optically by light scattering (see next section). As mentioned in the introduction, it was already found more than 100 years ago that nerves contract when they are excited. The contraction experiment was made because it was assumed that thickness changes in nerves might be small. This is indeed the case. Fig. \ref{iwasa_tasaki_a_b} (left) shows a well-known experiment by Iwasa and Tasaki \cite{Iwasa1980a}. They measured thickness changes in squid axons using a cantilever-method described in \cite{Iwasa1980b}. They found a deflection of the cantilever by about 1 nm, which corresponds to about 20\% of the thickness of a biological membrane. With the same setup one can also measure a force exerted on the cantilever, which for the squid axon can be up to 700 nN (two recordings in Fig. \ref{iwasa_tasaki_a_b}, right). Mechanical and electrical recordings are in phase.

\begin{figure}[htbp]
\centering
\includegraphics[width=225pt,height=118pt]{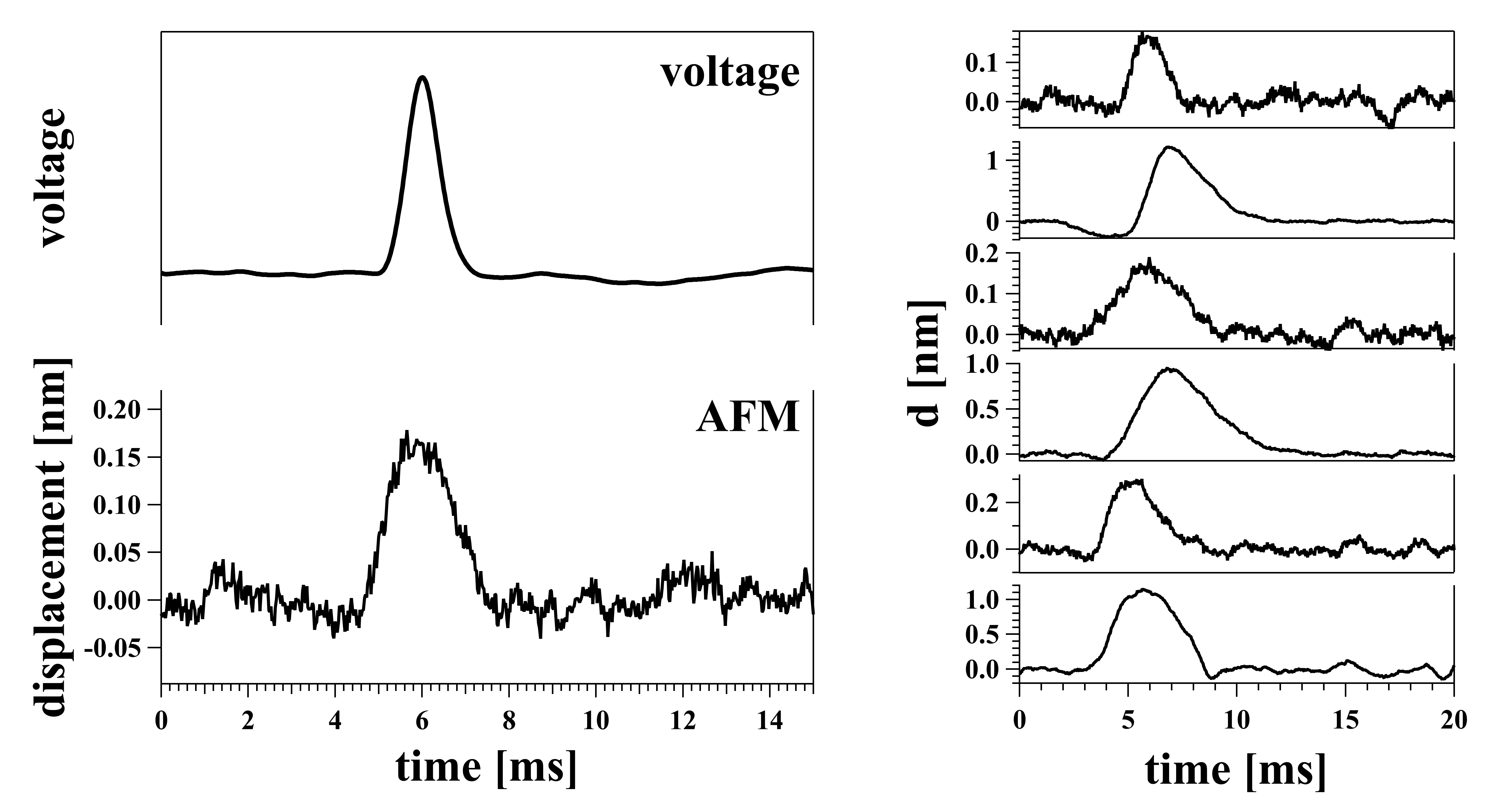}
\caption{Thickness change (bottom) of giant axons from lobster connectives during the action potential (top) measured by an atomic force microscope. Left: In this experiment the magnitude of the vertical dislocation is 0.15 nm. Right: The vertical dislocation in similar experiments on other giant axons ranges up to 1 nm, consistent with the data of Iwasa and Tasaki in Fig. \ref{iwasa_tasaki_a_b} (left). Adapted from \cite{GonzalezPerez2016} }
\label{gonzalezperez_2016_afm}
\end{figure}

Gonzalez-Perez et al. found similar changes in membrane thickness by atomic force microscopy \cite{GonzalezPerez2016}. They used single giant axons from lobster connectives and found a vertical dislocation of up to 1 nm at constant force, which was in phase and roughly of the same shape than the electrical recording. Fig. \ref{gonzalezperez_2016_afm} shows six recordings on different neurons. The magnitude of the signal varies due to the quality of the contact of the cantilever and the nerve surface. The upper bound of the recordings was always around 1 nm, similar to the magnitude of the changes of Iwasa and Tasaki in Fig. \ref{iwasa_tasaki_a_b} (left).

\begin{figure}[htbp]
\centering
\includegraphics[width=225pt,height=75pt]{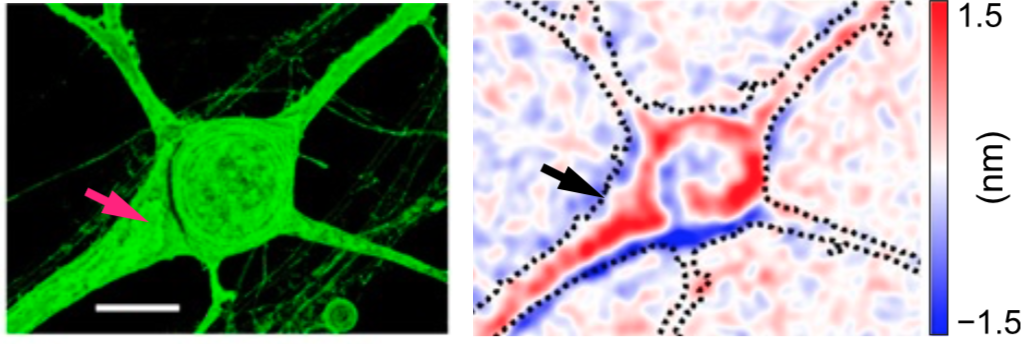}
\caption{Nanometer-scale membrane deformations from neurons from rat cortical tissues measured by high-speed interferometric imaging show thickness change of 1--2 nm. Left: Confocal microscopy image. Right: Interferometric image of the spike-induced deformation on the same nerve. From Ling et al., 2020 \cite{Ling2020} (PNAS license).}
\label{ling_deisseroth2020_interferometry}
\end{figure}

\begin{figure}[htbp]
\centering
\includegraphics[width=169pt,height=163pt]{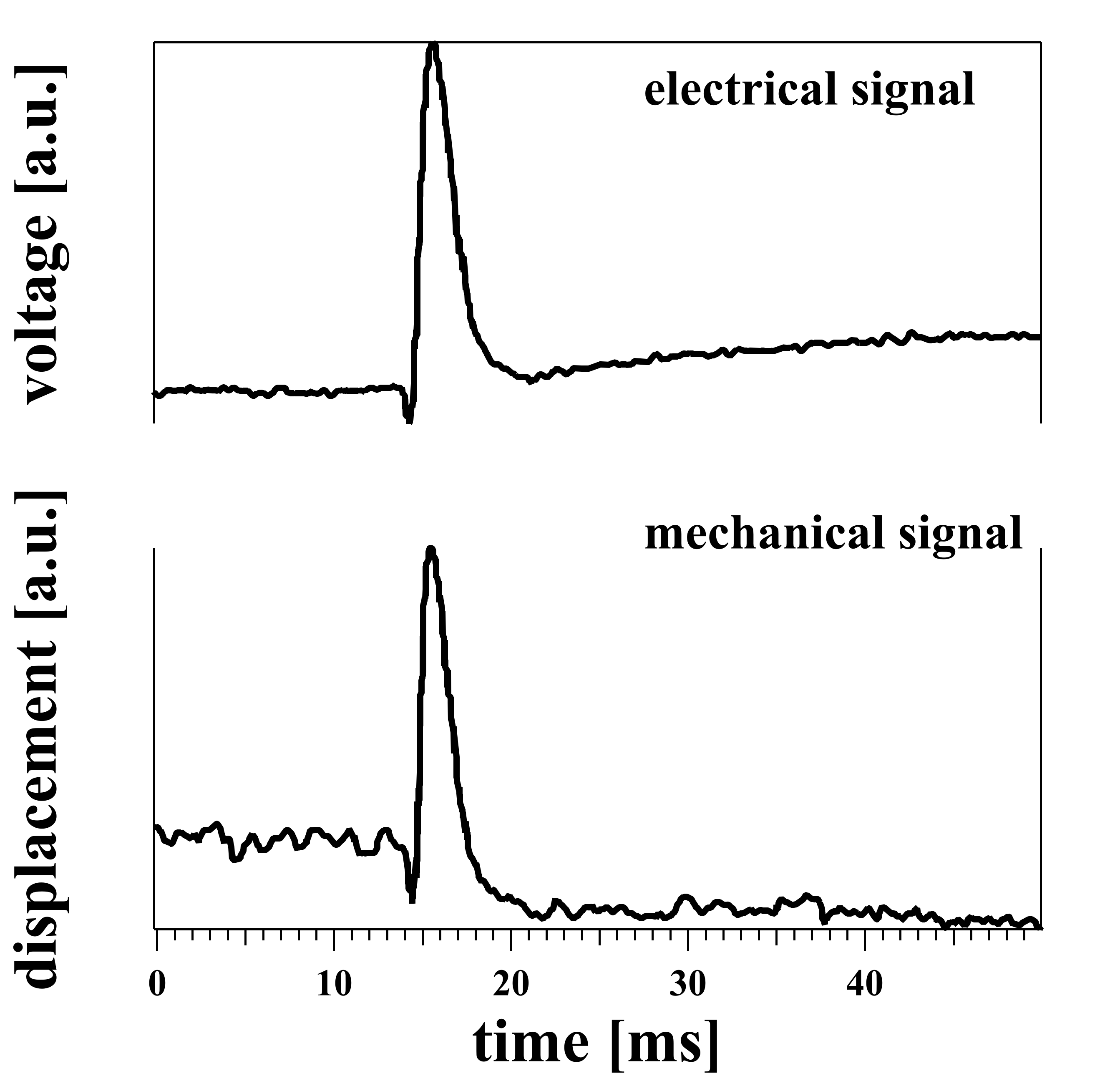}
\caption{Voltage and thickness change of nerve terminals of the mammalian neurohypophysis. The mechanical signal was measured by anatomic force microscope and the electrical signal by a fluorescence marker.Adapted from \cite{Kim2007}.}
\label{kim_salzberg_synapse_2007}
\end{figure}

Another very interesting technique is interferometry (high-speed quantitative phase imaging) described in Ling et al. \cite{Ling2020}. This technique has the advantage that one can follow thickness changes of membranes in a neural culture in wide-field images in a manner that is simultaneously time and space-resolved. In \cite{Ling2020} this is demonstrated in several movies. Fig. \ref{ling_deisseroth2020_interferometry} shows the height-profile in a complete neuron at one given point in time. One can recognize changes of the order of $\pm 1.5$ nm. Thus, in all of the above recordings, thickness changes of a neuron of the order of 1 nm were found. It should be pointed out the this change roughly corresponds to the change in thickness when one or two membranes goes through a melting transition (thickness change of about 17 \% \cite{Heimburg1998, GonzalezPerez2016} of a membrane thickness of about 5 nm). Thickness changes were also observed in synapses. Fig. \ref{kim_salzberg_synapse_2007} shows the voltage change and the thickness change of nerve terminals in the neurohypophysis.

In all recordings, electrical and mechanical changes were in phase and qualitatively nearly identical. This suggests, that the data represent different projections of the same phenomenon and should not be seen as independent events.

\subsection{Nerve contraction}
\label{nervecontraction}

\begin{figure}[htbp]
\centering
\includegraphics[width=224pt,height=149pt]{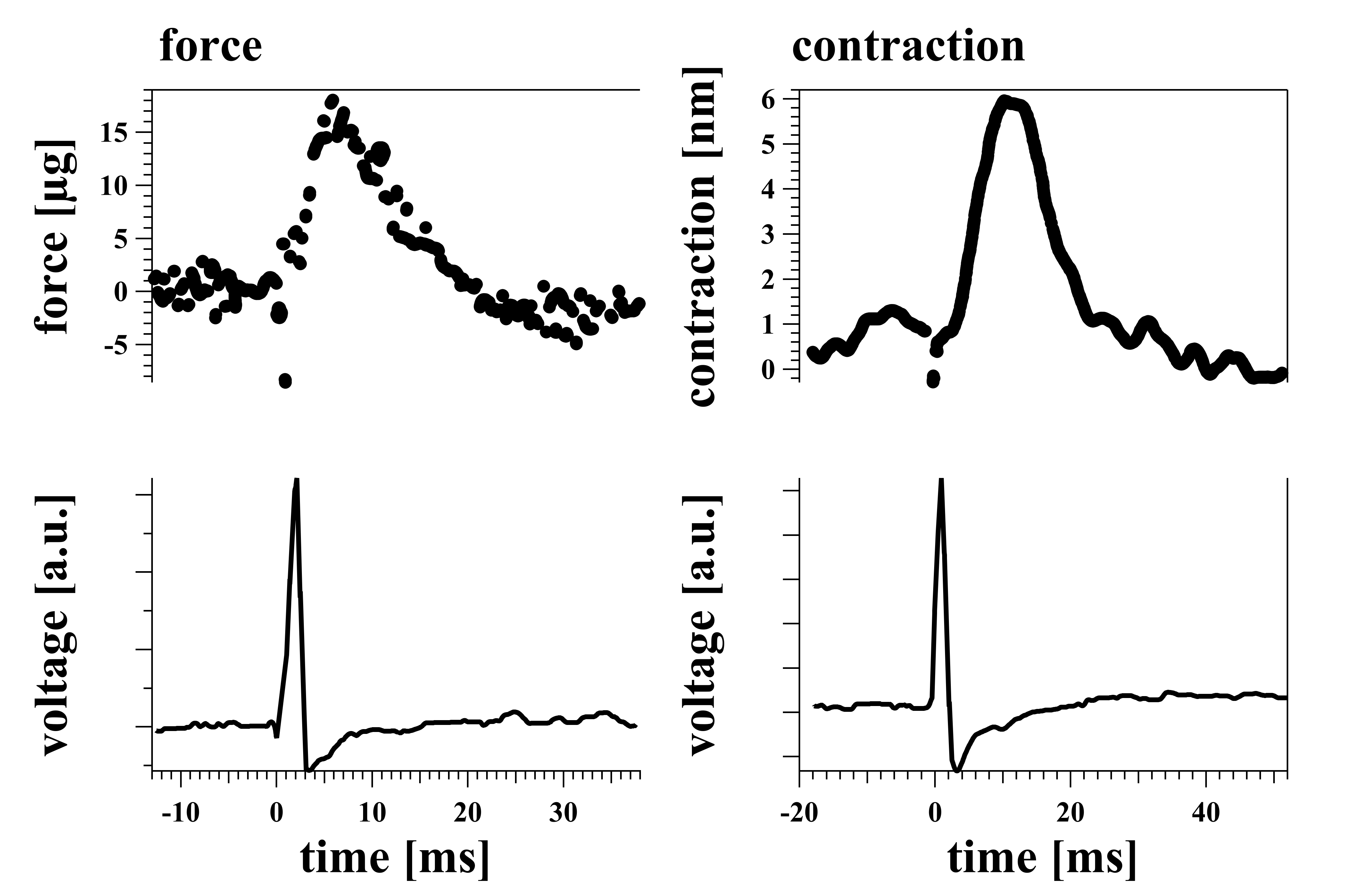}
\caption{Shortening of a crab nerve under the influence of a single action potential. Left: Force along the axon in units of a counter-weight. At maximum a force of 18 \textmu g $\equiv$ 177 nN was found. Right: The shortening of the nerve. Adapted from \cite{Tasaki1980b}.}
\label{tasaki_iwasa1980_contraction}
\end{figure}

As already suggested by \cite{Wilke1912a, Wilke1912b} (see introduction), nerves contract during the action potential. Since the action potential travels along the length of the axon, the time span of contraction is not equal to the time scale of the action potential but rather proportional to the dwell time of the action potential in the total axon. The duration of the contraction can be many times longer than the time scale of the action potential at a given position. This is shown in Fig. \ref{tasaki_iwasa1980_contraction}. Tasaki and Iwasa \cite{Tasaki1980b} used a cantilever to measure changes in length and the force applied to the cantilever (in units of a counterweight where 1 \textmu g $\equiv$ 9.81 nN). They found both a force of about 177 nN and a shortening of the axon of about 6 nm on a time scale much longer than the time scale of the action potential. The magnitude of these changes probably has to be taken with caution because nerves are very soft, they are under the load of their weight, and they possess inertia. Therefore, the magnitude of length changes is likely being underestimated.

\begin{figure}[htbp]
\centering
\includegraphics[width=225pt,height=104pt]{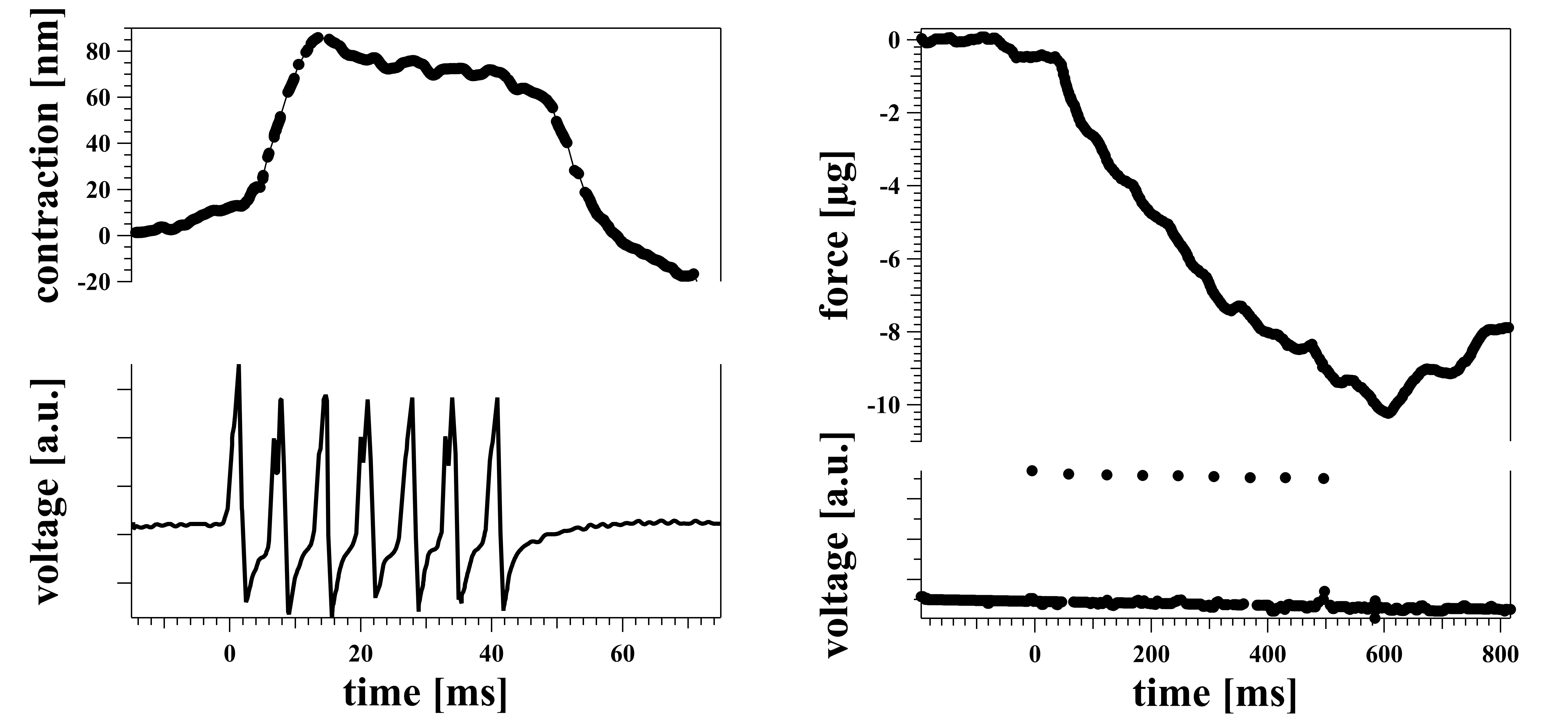}
\caption{Contraction of nerves during repetitive stimulation. Left: Records of tetanic contraction of 20-mm long desheathed crab nerve. The contraction lasts as long as there are action potentials in the nerve. Adapted from \cite{Tasaki1982a}. Right: Shortening of a garfish olfactory nerve associated with the generation of a train of nerve pulses. The top trace shows downward movements corresponding to the development of a force tending to pull the tip of a piezoelectric bender downwards. The maximum force is approximately 10 \textmu g $\equiv$ 98 nN. The lower traces (9 dots) indicate the nerve stimulations. Adapted from \cite{Tasaki1989}.}
\label{tasaki_byrne_1982_1989_contraction}
\end{figure}

Fig. \ref{tasaki_byrne_1982_1989_contraction} (left) shows that under repetitive stimulation of a desheathed crab nerve, the shortening of the nerve lasts as long as the pulse train is present \cite{Tasaki1982a}. A total length change of about 80 nm was observed, i.e., the contraction is much larger than the thickness change of the neurons described in the previous section. Olfactory nerves show very slow propagation velocities (see table \ref{tab:prop_nerv_1}). During repetitive stimulus, the length changes of each action potential accumulates in the axon. Fig. \ref{tasaki_byrne_1982_1989_contraction} (right) shows the accumulating increase of the force (measured in \textmu g in units of a counterweight) in a garfish olfactory nerve \cite{Tasaki1989}. A maximum force of 10 \textmu g $\equiv$ 98 nN was found. Both, force and contraction should be proportional to the number of pulses coexisting in the axon.

\subsection{Optical recordings}
\label{opticalrecordings}

\begin{figure}[htbp]
\centering
\includegraphics[width=225pt,height=101pt]{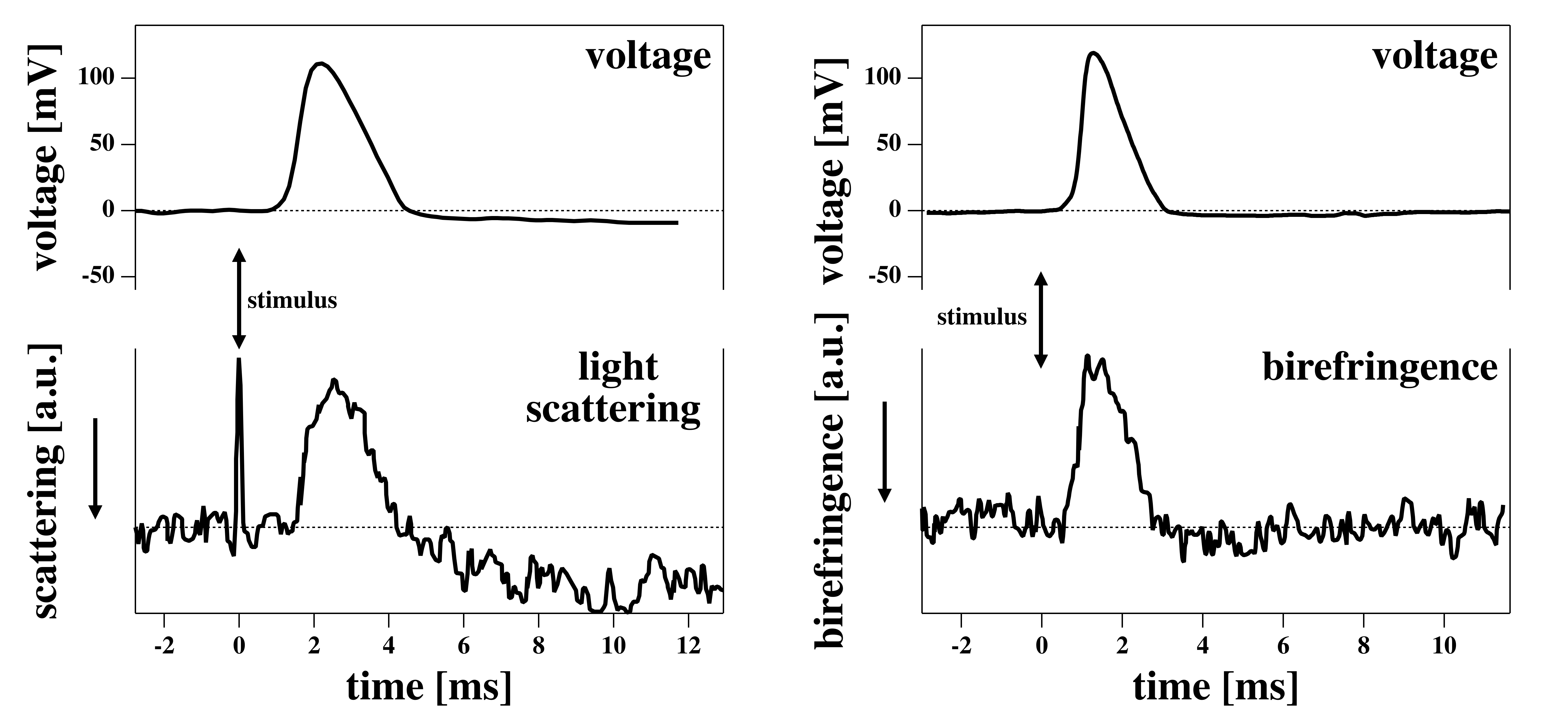}
\caption{Optical recordings on the squid axon. Left: Light scattering at 45$^\circ$ compared to voltage change. Right: Birefringence compared to voltage changes. Adapted from \cite{Cohen1968}.}
\label{cohen1968_scattering_birefringence}
\end{figure}

It has long been known that geometrical changes in nerves can be observed by light scattering. Fig. \ref{cohen1968_scattering_birefringence} (left) shows a recording of white light scattering at an 45$^\circ$ angle during the action potential of in a squid axon \cite{Cohen1968}. Light scattering monitors changes in geometry, in refractive index, or both. The time course is the same as the electrical recording (upper trace). Further evidence for light scattering changes during the action potential was reported by Tasaki and collaborators \cite{Tasaki1968} for spider crab nerves, squid nerves and lobster leg nerves. Light scattering changes were also reported for neurons of aplysia (a large sea slug) \cite{Stepnoski1991}. The authors concluded that \emph{``An analysis of the data indicates that the radial component of the index of refraction of the membrane increases and the tangential components decrease concomitant with an increase in membrane potential. This is suggestive of a rapid reorientation of dipoles in the membrane during an action potential.''} It is consistent with the hypothesis of \cite{Wei1972} on the coupling of a dipolar reorientation of membrane molecules and reversible heat release in nerve. One possible explanation for such dipolar reorientation is a change in the phase state of the lipid membrane because it is known that the dipole potential of lipid monolayers is different in gel and fluid state of the membrane \cite{Vogel1988, Beitinger1989, Brockman1994, Lee1995}.

\begin{figure}[htbp]
\centering
\includegraphics[width=225pt,height=103pt]{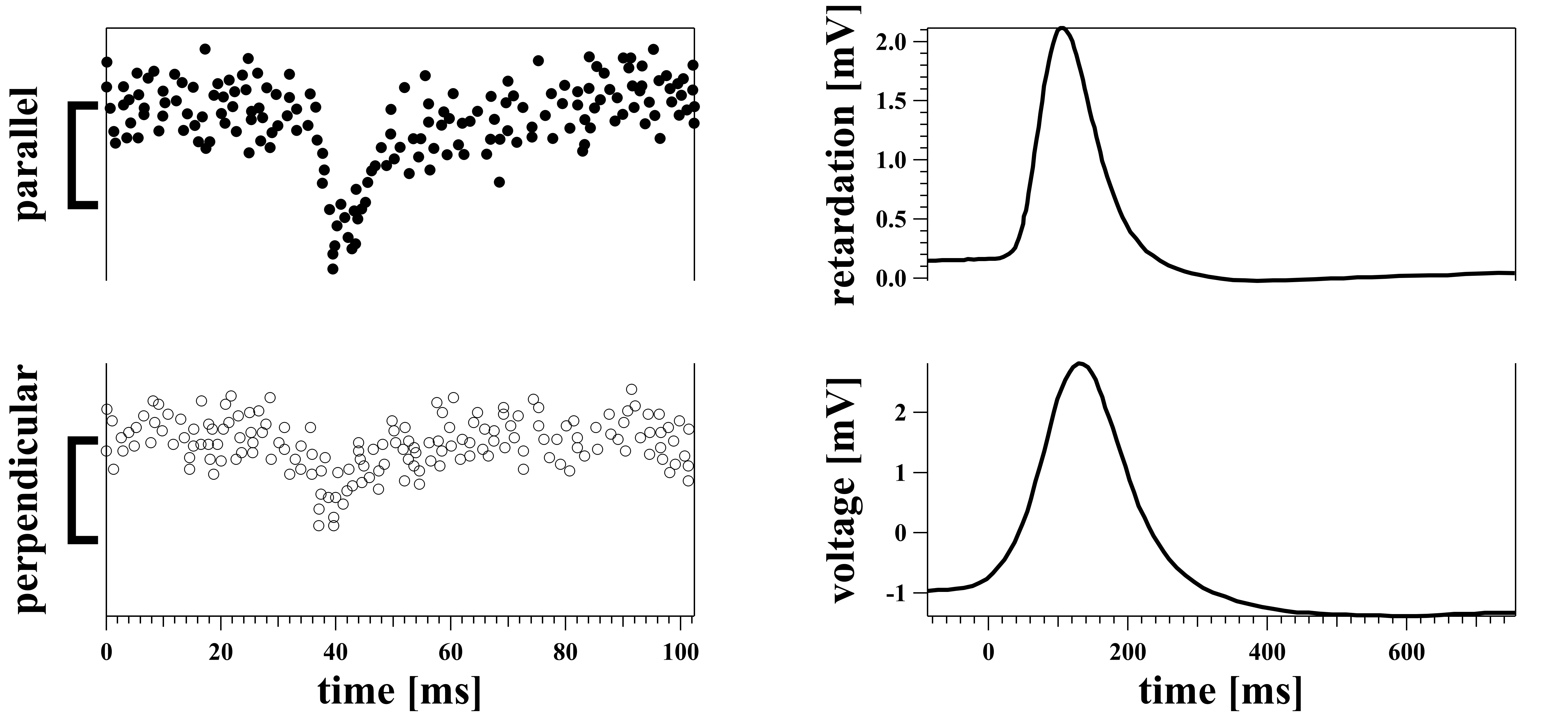}
\caption{Left: Changes in fluorescence polarization during excitation in a crab nerve stained with Pyronin B. The bottom record was taken under the same experimental conditions except that the analyzer was rotated by 90$^\circ$. Adapted from \cite{Kobatake1971}. Right: Birefringence changes during the action potential of pike olfactory nerve. Upper curve: Retardation changes. Lower curve: action potential. Adapted from \cite{vonMuralt1976}.}
\label{vonmuralt_retardation_kobatake_fluorescence}
\end{figure}

The right hand panel of Fig. \ref{cohen1968_scattering_birefringence} shows birefringence changes during the action potential. Birefringence is a material property that indicates that the refractive index is different for different polarizations of light. During a birefringence recording the relative changes of the intensity of the light scattering with different polarization is recorded. Fig. \ref{cohen1968_scattering_birefringence} (right) shows that one finds birefringence changes that are proportional to the action potential. The authors attributed these changes partially to a reorientation of elongated proteins that are tangential to the membrane surface and the reorientation of lipids of the Schwann cells and the axon. The authors also reported similar birefringence changes in crab nerves. Changes in birefringence in crab spider nerves and squid axon were also reported by \cite{Tasaki1968}. These changes were attributed to retardation, i.e. a relative change of the intensities of different polarizations leading to a rotation in optical axis \cite{Cohen1970}. As mentioned, such optical activity could originate from the rotation of proteins and lipids, and are most likely properties of the nerve membrane. Changes in retardation during the action potentials in pike olfactory nerve were reported by \cite{vonMuralt1976}, see Fig. \ref{vonmuralt_retardation_kobatake_fluorescence} (right). Kobatake and collaborators \cite{Kobatake1971} showed that there are changes in fluorescence polarization of a lipid dye during the action potential (see Fig. \ref{vonmuralt_retardation_kobatake_fluorescence}, left), indicating that the lipid membrane is at least partially involved in the changes of the optical properties.

One can summarize the optical properties as follows: The optical changes are in phase with the action potential and with the mechanical changes. They must be due to geometry changes and\slash or to changes in refractive index and molecular orientations. Thus, they indicate structural changes of the excited part of the neurons both on macroscopic and on molecular level. In particular, the changes in orientation of molecules from tangential to normal to the membrane are consistent with changes in the head group orientation of lipids. Changes in lipid order are also found when measuring the fluorescence polarization of lipid dyes. These changes suggest changes in the physical state (for instance a fluid-gel transition) of membrane molecules including the membrane lipids.

Summarizing, one can conclude that both electrical, mechanical, optical and thermal changes in the nerve membrane are in phase and most likely represent the same phenomenon which appears as a more general thermodynamic phenomenon.


\section{Theoretical considerations}
\label{theoreticalconsiderations}


\subsection{The soliton theory}
\label{thesolitontheory}

\begin{figure}[htbp]
\centering
\includegraphics[width=225pt,height=63pt]{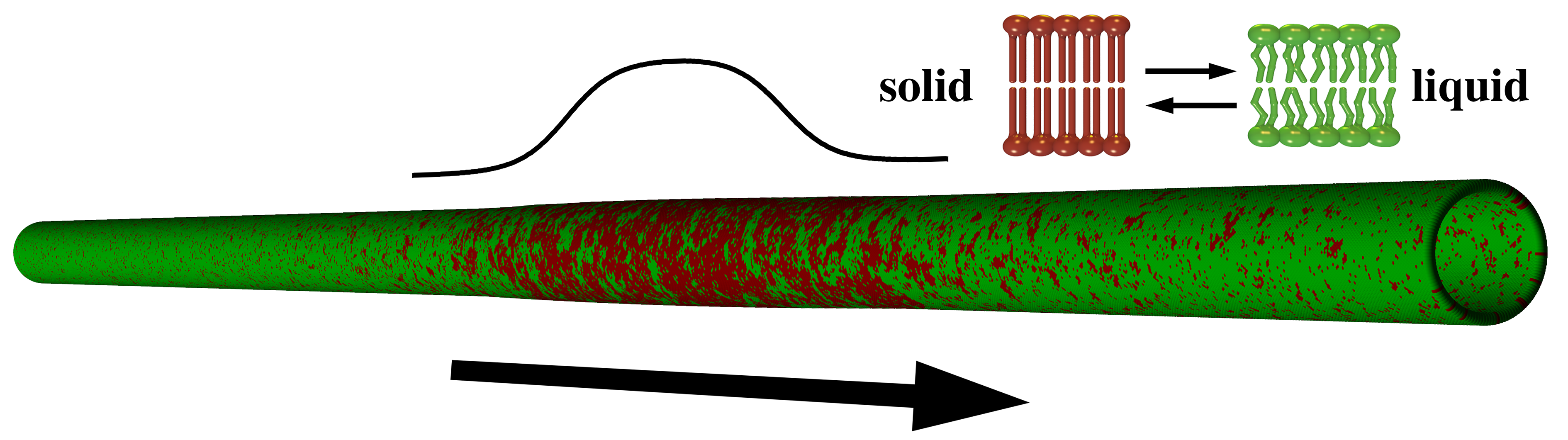}
\caption{Schematic representation of a soliton traveling in an axon. Red (solid = gel phase) and green (liquid = fluid phase) color denotes changes in the physical state of the membrane under the influence of the action potential. The soliton travels with a velocity slightly lower than the speed of sound. From \cite{GonzalezPerez2016}.}
\label{soliton1}
\end{figure}

The soliton theory is based on the hydrodynamics of a biological membrane. It relates the propagation of the nervous impulse to an electromechanical soliton in the membrane, which is coupled to a change in the phase state of the lipid membrane. A phase transition from a solid (gel) to a liquid (fluid) state can be found in many biological membranes about 10--15$^\circ$ below physiological temperature \cite{Muzic2019, Fedosejevs2022, Faerber2022}. It is responsible for a nonlinear dependence of the elastic constants on temperature, pressure and other intensive variables. If a membrane is compressed from a fluid state at physiological conditions, it is moved into the transition range. As a consequence, the elastic constants of the membrane change, heat is released, the membrane increases its thickness and decreases its area. In particular, the sound velocity approaches a minimum, which has important consequences for the wave equation. Sound propagation in such a non-linear membrane leads to the possibility of soliton propagation \cite{Heimburg2005c, Lautrup2011}, i.e., a localized sound pulse. A soliton traveling in a cylindrical membrane is schematically shown in Fig. \ref{soliton1}. It shares remarkably many properties with the traveling nervous impulse.

\begin{figure}[htbp]
\centering
\includegraphics[width=169pt,height=134pt]{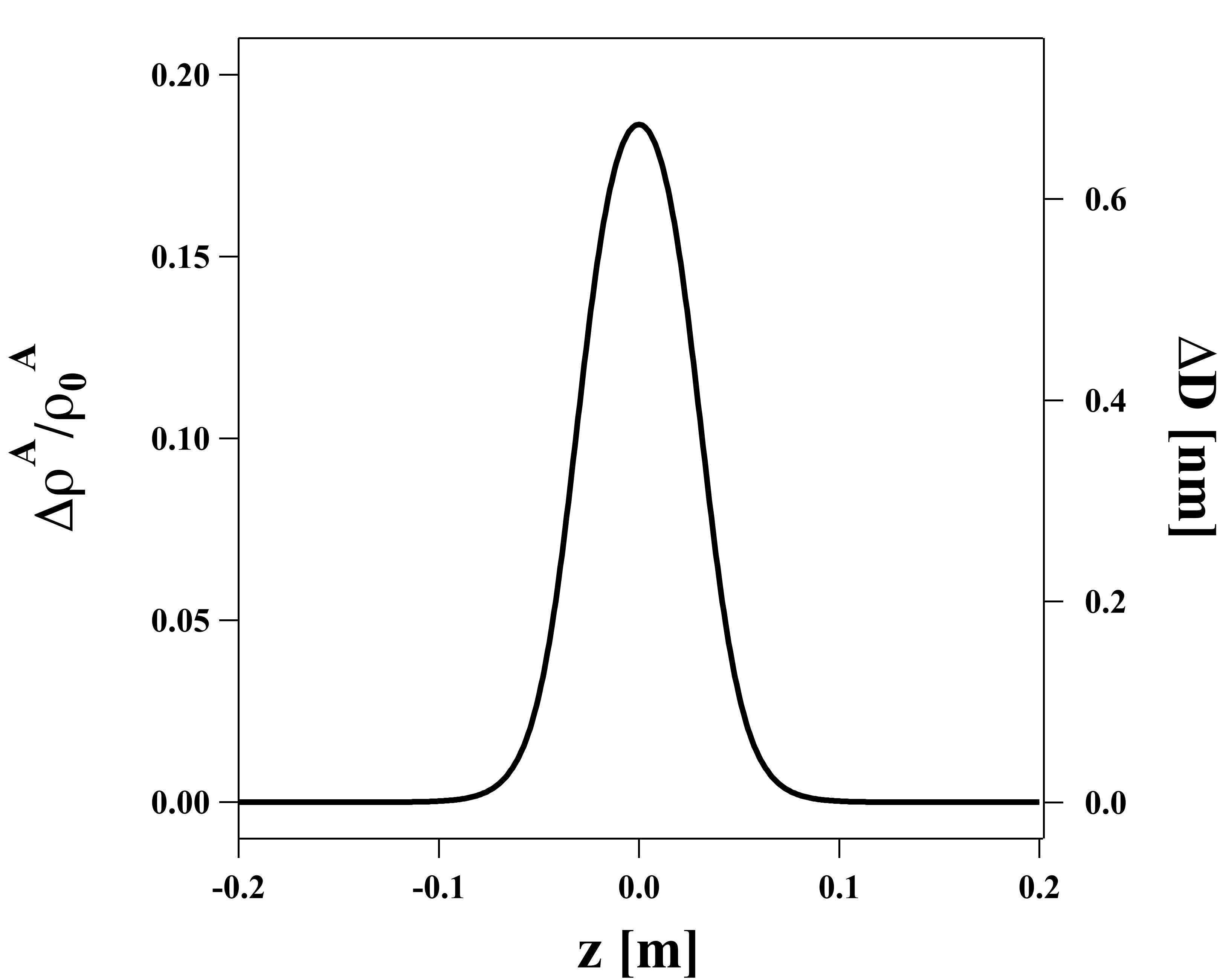}
\caption{Soliton calculated for the thermodynamic parameters of DPPC LUV at 45$^\circ$C at a velocity of $v=0.655\, c_0 \approx 115$ m\slash s (see \cite{Heimburg2005c} for details). The coordinate $z=x-vt$ denotes the width of a soliton in a frame moving with the velocity $v$ of the soliton. The relative area density change at maximum is approx. 19\%. The thickness change at maximum is around 0.8 nm.}
\label{soliton_v0655}
\end{figure}

The soliton is the solution of the differential equation
\begin{equation}\label{eq:soliton01}
	\frac{\partial^2 \Delta \rho^A}{\partial t^2}=\frac{\partial}{\partial x}\left[\left(c_0^2+p \Delta \rho^A+q(\Delta \rho^A)^2\right)\frac{\partial \Delta \rho^A}{\partial x}\right]-h\cdot \frac{\partial^4 \rho^A}{\partial x^4}\;,
\end{equation}
where $\Delta \rho^A$ is the change in lateral mass density of the membrane, $c_0$ is the 2-dimensional sound velocity of the membrane at physiological temperature, $p$ and $q$ are empirical coefficients that describe the shape of the melting transition in the membrane as a function of density, and $h$ is a dispersion parameter that describes the frequency-dependence of the sound velocity (see \cite{Heimburg2005c} for details). The dispersion parameter $h$ determines the pulse width. A typical soliton is shown in Fig. \ref{soliton_v0655}. The horizontal coordinate $z$ indicates a distance in a coordinate system moving with the velocity $v$ of the soliton, $z=x-vt$. The amplitude of the soliton depends on its velocity. Slower solitons display a larger amplitude. For the thermodynamic parameters of DPPC\footnote{dipalmitoyl phosphatidylcholine} (given in \cite{Heimburg2005c}), the maximum amplitude from theory is $\Delta \rho^A/\rho_0^A=0.212$ which is approximately the relative area-density change for a change from the fluid to the gel state of a membrane (0.227 in DPPC at $T_m$). The maximum amplitude corresponds to a minimum soliton velocity of $\approx 0.65\cdot c_0$, where $c_0=176$ m\slash s is the calculated 2D sound velocity in a DPPC membrane. This implies that the soliton moves with the velocity typical for myelinated motor neurons. The soliton consists of a region of ordered membrane traveling in a disordered fluid (Fig. \ref{soliton1}).

Since the volume change of a membrane in the transition from fluid to gel is just about -4\%, a change of the thickness $D$ of the membrane of roughly 15.5\% accompanies the area changes \cite{Heimburg1998}, which corresponds to 0.6 nm for DPPC. Thus, the change of the thickness of a cylinder with two opposing membranes is expected to be about 1.2 nm for the maximum amplitude soliton. This is very close to the thickness changes of $\approx$ 1 - 1.5 nm observed experimentally (see section \ref{mechanicalrecordingsoftheactionpotential:thicknesschanges}).

The biological membrane is a capacitor with $C_m=\epsilon_o \epsilon \cdot A/D$ ($\epsilon_0$ is the vacuum permittivity, $\epsilon$ is the dielectric constant, $A$ is the membrane area and $D$ is the membrane thickness). The order of magnitude of the capacitance in biomembranes is 1 \textmu F\slash cm$^2$, which can also be calculated from the membrane dimensions given above and a dielectric constant $\epsilon\approx 4$ of the membrane core \cite{Heimburg2012}. A change from a fluid to a gel membrane reduces the capacitance by about a factor of 1.5 to 2 \cite{Heimburg2012, Zecchi2017}. A soliton will therefore transiently reduce the capacitance of the membrane. It might also change the polarization of the membrane because the head groups of the gel and the fluid membrane membrane display different orientation. Thus, the soliton must display an electrical component in phase with the density change and one expects both capacitive currents and voltage changes. The soliton can therefore be considered a piezoelectric or an electromechanical pulse.

Due to the higher area density in the region of the soliton, the nerve axon contracts (as seen experimentally in nerves, see section \ref{nervecontraction}). The maximum amplitude solution has a velocity of about 100 m\slash s \cite{Heimburg2005c}, similar to the propagation velocity of myelinated motor-neurons. Taking the coupling to water into account (which adds to the inertial mass of the membrane), results in propagation velocities of the order of a few meters per second, similar to the lower propagation velocity in non-myelinated nerves and neurons (see section \ref{velocityandlengthofthenervepulse}) and density pulses in lipid monolayers \cite{Kappler2017, Schneider2021}. A soliton is naturally accompanied by a reversible release of heat into its environment, which corresponds to the latent heat of the melting transition. As described in the introduction, such reversible heat changes have been reported for nerves.

Summarizing, the soliton theory captures many properties of the nerve pulse qualitatively, and quite a few quantitatively.


\subsection{Nerve contraction in the soliton theory}
\label{nervecontractioninthesolitontheory}

\begin{figure}[htbp]
\centering
\includegraphics[width=225pt,height=86pt]{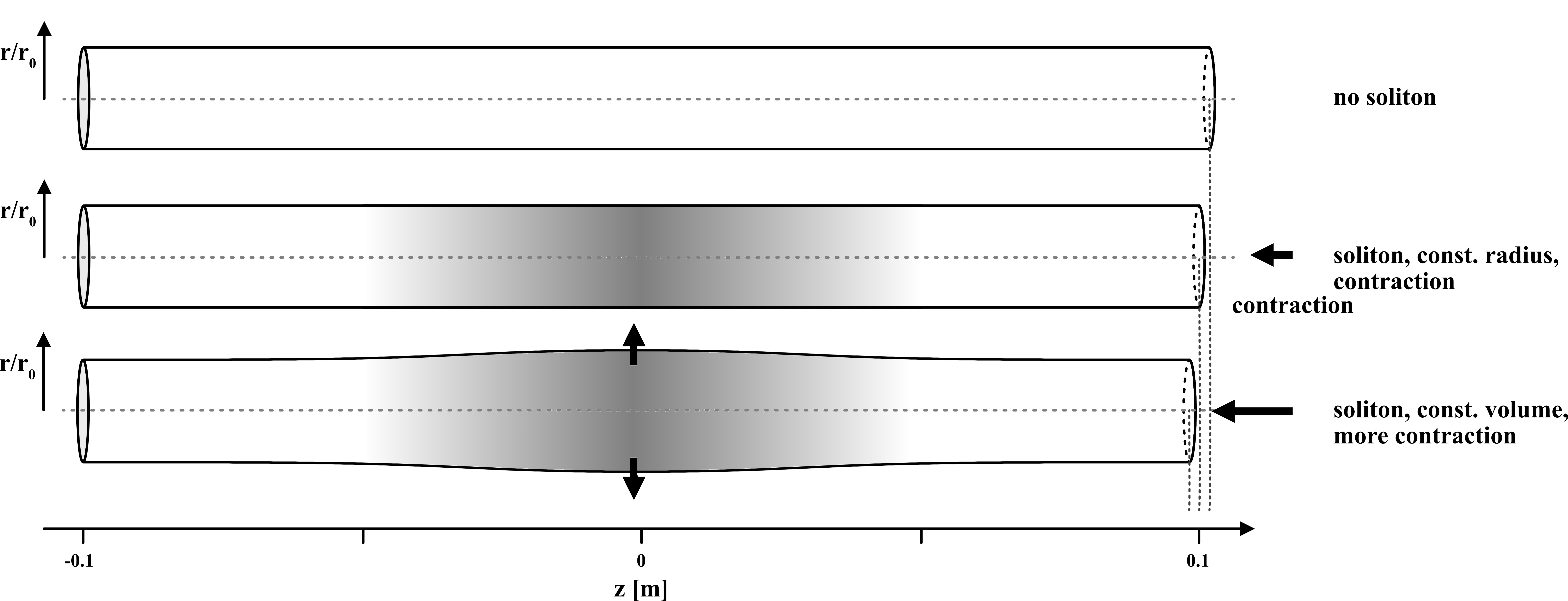}
\caption{Geometric changes in the soliton theory. Top: No soliton. Middle: If the radius of the axon is constant, the contraction is proportional to the area change. Bottom: If the volume of the axon is constant, the soliton will lead to a peristaltic pulse with pronounced changes in thickness and a larger contraction of the nerve as in the middle panel. }
\label{nerve_contraction}
\end{figure}

The soliton consists of a region on the axon in which the area mass density is higher. As a consequence, the membrane contracts in the presence of a soliton, as shown in Fig.\ref{nerve_contraction}. Both the surface area of the axon and the area-to-volume ratio change. In the following, we will consider two limiting cases: 1. The membrane is impermeable and the aqueous volume of the axon stays constant. 2. The membrane is permeable for the electrolyte and the radius of the axon is assumed constant. The soliton consists of a region of membrane moved into the phase transition, where membranes become leaky to ions and water \cite{Blicher2009, Heimburg2010, Mosgaard2013b} and in such a scenario water would flow through the membrane?

\vspace{0.3 cm}
\textbf{Constant volume:}
Let us consider a water filled cylinder surrounded by a membrane with uniform area density. It shall have open ends. Its surface area is $A_0=2\pi r_0\cdot x_0$, where $r_0$ is the radius of the cylinder and $x_0$ is its length. The volume is given by $V_0=\pi r_0^2\cdot x_0$. Thus,
\begin{equation}\label{eq:theor_contraction01}
r_0=\sqrt{\frac{V_0}{\pi x_0}}\quad\mbox{and}\quad A_0=\sqrt{4\pi V_0 x_0}\;.
\end{equation}
If the surface area of the cylinder contracts ($A<A_0$), its length contracts to a length $x<x_0$. We assume that the walls of the cylinder are impermeable to water and therefore the volume stays constant, $V=V_0$. Thus,
\begin{equation}\label{eq:theor_contraction02}
r=\sqrt{\frac{V_0}{\pi x}}\quad\mbox{and}\quad A=\sqrt{4\pi V_0 x} \quad\Rightarrow\quad \frac{A}{A_0}=\sqrt{\frac{x}{x_0}}
\end{equation}
and
\begin{equation}\label{eq:theor_contraction03}
\frac{r}{r_0} =\frac{A_0}{A}=\frac{\rho^A}{\rho_0^A} \;,
\end{equation}
where $\rho^A$ is the area density of the membrane. We see that the relative change in radius is equal to the relative change in area density, and the relative change in length is proportional to the square root of the relative area change.

Let us now assume that the nerve axon consists of many cylindrical segments with infinitesimal length $dx$. We make the simplification that no aqueous medium flows along the axon in order to obtain an approximation for the changes in axon radius under the influence of the action potential. Now we can calculate the change in radius as a function of position, and obtain the total contraction of the axon by integrating of the infinitesimal length changes. This is shown in Fig. \ref{nerve_contraction} (bottom). One finds a peristaltic pulse with changes in radius and an overall contraction. For a demonstration we take the soliton from Fig. \ref{soliton_v0655}. It has a maximum relative density change of $\Delta \rho^A/\rho_0^A=0.186$, corresponding to $\rho^A/\rho_0^A=1.186$. The width of the soliton at half height is 6.56 cm. We obtain a change in length of the total axon of $2.58$ cm, corresponding to about 38 \% of the soliton width. This change only depends on the width of the soliton and is independent of the total length of the axon (which could be of the order of one meter in a human motor neuron).

\vspace{0.3 cm}
\textbf{Constant radius:}
As shown by \cite{Papahadjopoulos1973, Sabra1996, Blicher2009}, membranes are permeable to ions, water and small molecules in their melting transition. Thus, the boundary condition of constant volume may not apply. In fact, even though the axon contracts, its radius does not seem to change by more than 1--1.5 nm (see section \ref{mechanicalrecordingsoftheactionpotential:thicknesschanges}) indicating that the radius of the axon is nearly constant in a living nerve.

Let us again consider a cylindrical element. If the membrane contracts (i.e., it increases its area density, $\rho^A$) under conditions of constant radius, its volume change must be proportional to the area change,
\begin{equation}\label{eq:theor_contraction04}
\frac{V}{V_0} =\frac{A}{A_0}=\frac{x}{x_0}=\frac{\rho_0^A}{\rho^A} \;.
\end{equation}
If the axon consists of infinitesimal cylindrical elements of length $dx$, one can obtain the changes in length from the area density profile of a soliton by integration. This is shown in Fig. \ref{nerve_contraction} (middle). For the soliton in Fig. \ref{soliton_v0655}, we find a contraction of the axon by 1.44 cm, corresponding to a change of about 20\% of the width of the soliton. This is about equal to the change of the area of a membrane undergoing a change from the fluid to the gel state. The change in cylinder length is proportional to the relative change in area. Therefore, the contraction in the constant radius case is smaller than in the case of constant volume, where the length change is quadratic in the relative change in area. Since in the constant radius case the volume of the axon decreases proportional to the change in length when a soliton passes, there must be flow of electrolyte out and back into the nerve. We describe this in more detail in the discussion section.

Summarizing, we see that the contraction of a motor neuron with free ends (conduction velocity \textsubscript{100} m\slash s) as described by the soliton theory is about 1.4 to 2.6 cm depending on the boundary condition. For systems with a conduction velocity of 1 m\slash s, one expects contractions on the order of 140--260 \textmu m.


\subsection{Dependence of excitation threshold on stretch}
\label{dependenceofexcitationthresholdonstretch}

\begin{figure}[htbp]
\centering
\includegraphics[width=225pt,height=76pt]{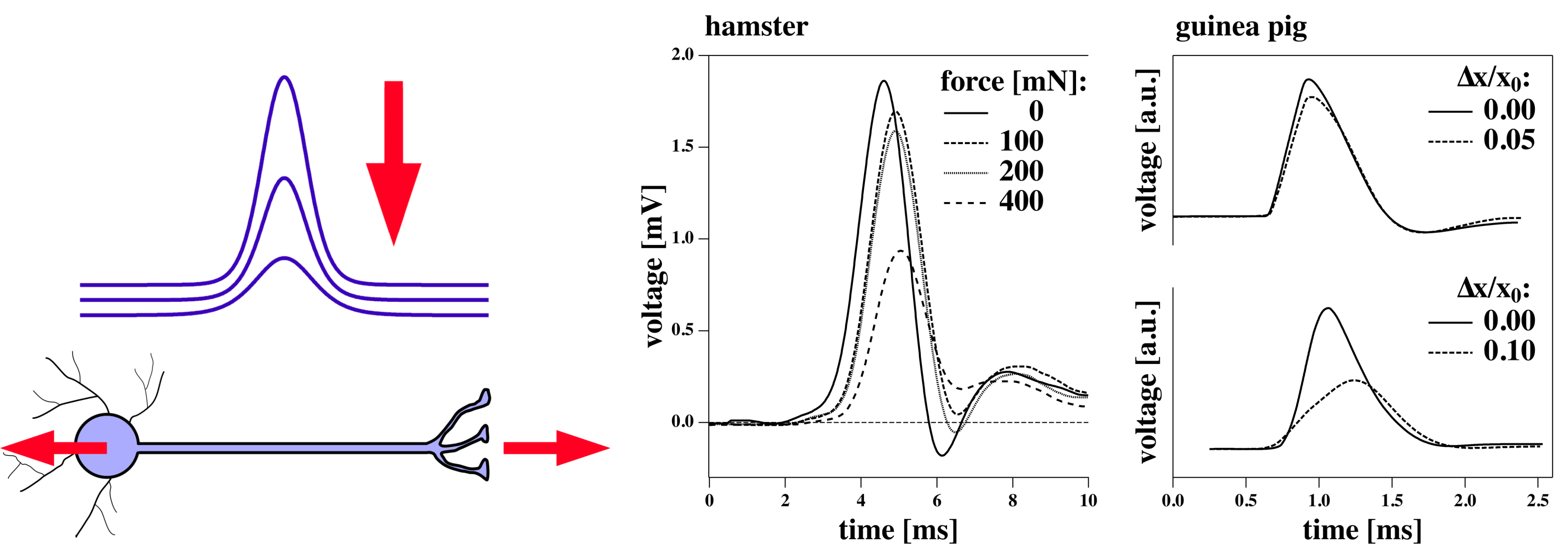}
\caption{Consequences of nerve stretching. Left: Schematic representation of the reduction of the amplitude of an action potential upon stretch as determined from the soliton theory. Center: Reduction in action potential amplitude upon stretch of hamster sciatic nerve at four different forces, 0, 100, 200 and 400 mN. Adapted from \cite{Stecker2011}. Right: Reduction of the compound action potential amplitude of guinea pig sciatic nerve at three different elongations: $\Delta x/x_0 =$ 0, 0.05 and 0.10. Adapted from \cite{Li2007}. Figure from \cite{Heimburg2022a}.}
\label{stretch_heimburg2022}
\end{figure}

In the soliton theory, the distance between the melting temperature and physiological temperature is proportional to the stimulation threshold because the free energy difference between the solid membrane and the fluid membrane increases \cite{Wang2018}. The melting transition in biological membranes depends on the lateral pressure because
\begin{equation}\label{eq:theor_stretch}
T_m=\frac{\Delta E+\Pi \cdot\Delta A}{\Delta S} \;,
\end{equation}
where $\Pi$ is the lateral pressure and $\Delta S$ is the melting entropy. Tension in the membrane corresponds to a negative pressure. This implies that upon an increase of the tension in a neuron, the melting transition is moved to a lower temperature, further away from the physiological temperature. This results in a larger threshold for nerve stimulation. This was shown in detail in \cite{Wang2018}. Therefore the prediction is that upon stretch of nerves, the threshold of nerve pulse excitation increases, or at constant stimulus that the amplitude of the nerve pulse decreases both when changing the force in the axon or when increasing its length. This was demonstrated in \cite{Heimburg2022a}. Fig. \ref{stretch_heimburg2022} shows how tension in the axon and extending its length leads to the reduction in the amplitude of compound action potentials in hamster and guinea pig sciatic nerves. This implies that mechanical force has a measurable effect on the excitability of nerves which would be impossible without a mechanical component of the action potential, i.e., if the nerve pulse was not accompanied by a contraction.

\section{Discussion}
\label{discussion}

In this paper, we discuss the experimental evidence for the mechanical nature of action potentials. As shown in section \ref{experimentalevidenceformechanicalchangesinnerves}, mechanical changes in thickness and length have been found in force measurements by AFM and similar methods, but also by light scattering and interference microscopy imaging experiments. We showed that the nerve pulse is macroscopic and that the action potential is larger than synaptic gaps, the internode-distance in myelinated nerves and even larger than nerve cells as a whole. In fluorescence and birefringence experiments it has been found that the polarization, the orientation and therefore probably the physical state of molecules in the membrane changes during the nerve pulse. This suggests that the nerve pulse expresses itself as a structural change both on macroscopic and microscopic scales.

Such considerations are not new but are nearly as old as the investigation of the electrical properties of nerves. It has been known since the mid-nineteenth century that nerves can be excited mechanically \cite{Tigerstedt1880, Rosenblueth1953, Yamada1961, Julian1962, Terakawa1982}. Due to the absence of any measurable net heat dissipation during nerve pulse conduction, it was suggested more than 100 years ago that the nerve pulse might consist of a reversible physical phenomenon \cite{Hill1912, Bayliss1915} similar to waves or sound. The present picture of nerve pulse conduction is the Hodgkin-Huxley model \cite{Hodgkin1952b} that focusses on dissipative electrical phenomena. It is inconsistent with the reversible heat change \cite{Hodgkin1964, Heimburg2021} and it does not consider mechanical phenomena explicitly. Mechanical properties can only enter in a non-thermodynamic manner by postulating mechanosensitive channel proteins (e.g., \cite{Coste2010, Coste2012}) that influence the membrane conductance. But it is unlikely that a structural change of a few channel proteins can account for the experimentally observed macroscopic changes in thickness and length.

The soliton theory \cite{Heimburg2005c} describes the nerve pulse as an adiabatic wave reminiscent of sound in which thermal, electrical and mechanical changes are coupled in an unambiguous manner. The adiabaticity of the nerve pulse is suggested by the reversible heat release during the action potential that was found experimentally in numerous nerves (reviewed in \cite{Heimburg2021}, see introduction). As we show here, the nerve pulse is a macroscopic event, and the spatial length of the nerve pulse can be on scales from many millimeters up to many centimeters, which is up to eight orders of magnitude larger than the size of single channel molecules. This justifies the use of macroscopic thermodynamic theory. As a parameter, the soliton theory contains only the speed of sound on macroscopic scales close to the melting transition of a membrane. The main variable is the mass density of the membrane. In a macroscopic theory, no knowledge of the properties of single molecules is required, but the macroscopic susceptibilities such as heat capacity or compressibility intrinsically depend on the chemical potentials of such molecules. The susceptibilities are dominated by thermal fluctuations on scales larger than single proteins and lipids \cite{Heimburg2005c}. In the soliton theory, the nerve membrane as a whole is temperature-, voltage- and mechano-sensitive.

The soliton theory has a high predictive power. This is a consequence of its thermodynamic nature where all coupling between different variables are strict. Numerous predictions of biological and medical relevance can be made, e.g., about essential tremor, the effects of membrane-soluble drugs, the action of lithium, and about fever and inflammation that were discussed in \cite{Heimburg2022b}. The soliton theory provides an explanation for anesthesia, because anesthetic molecules lower melting transitions via the freezing-point depression law and thus increases the stimulation threshold \cite{Heimburg2007c, Graesboll2014, Wang2018}. Hydrostatic pressure has an opposing effect because it increases transition temperatures. This leads to the well-known phenomenon of the pressure reversal of anesthesia that can quantitatively be calculated from the known pressure-dependence of transition temperatures \cite{Johnson1950, Johnson1970, Heimburg2007c}. This effect requires the presence of mechanical components of the nerve pulse because hydrostatic pressure couples to volume changes in a similar manner than lateral pressure couples to area changes. Since the soliton theory is derived from isentropic hydrodynamics, it suggests that due to momentum and energy conservation two colliding pulses pass through each other, as it is usually the case for wave-like phenomena. This has in fact been found in worm axons, lobster nerves and single lobster axons \cite{GonzalezPerez2014, GonzalezPerez2016, Wang2017}. The soliton consists of a region of an ordered lipid in a fluid environment. Qualitatively, the reversible heat release during the action potential (described in the introduction) is consistent with the latent heat of the melting transition of the lipids. During the nerve pulse, one finds thickness changes in the nerve on the order of 1--2 nm. This corresponds to the thickness change of two opposing nerve membranes of an axon during a phase transition from fluid to gel, which is the central element in the soliton theory. The soliton theory also predicts a contraction of nerves because a solitary pulse consists of a region of larger area density. The contraction predicted here is of the order of 20 \% of the width of the pulse, which could be of the order of a few 100 \textmu m up to centimeters when the thermodynamic parameters of the single lipid DPPC are taken into account - which probably overestimates the magnitude of the contraction in nerve membranes. There exist very few experiments that quantify the contraction of nerves. Tasaki and collaborators \cite{Tasaki1982a, Tasaki1989} found that crab nerves can contract by 80 nm upon stimulation. This is much less than what the soliton theory predicts but it is qualitatively in the same direction. Since nerves are very soft, and also carry a weight in the contraction experiment, the magnitude of the contraction is most likely underestimated. It would be of high interest to obtain more solid experimental numbers for the magnitude of contraction. If a neuron or a nerve is stretched, tension is created in the axon. Since the phase transition in the lipid membrane depends not only on temperature but also on hydrostatic and lateral pressure, tension in the membrane lowers the transition temperature and as a consequence will make a neuron less excitable. This was in fact found in various experiments that are discussed in \cite{Heimburg2022a} (see section \ref{dependenceofexcitationthresholdonstretch}). If neurons are fixed at the ends such that they cannot contract, the properties of pulses change and one finds pulse trains \cite{Villagran2011}. Thus, in neurons of fixed length, periodic pulses, depolarizations, and refractory periods emerge and are a consequence of mechanical changes and of mass conservation.

Since nerves can be stimulated mechanically, it may not be surprising that the brain can also be excited by ultrasound. In focused ultrasound neurostimulation (FUN), ultrasound is focused by a parabolic array of ultrasound transducers on a small region in the brain. For instance, Kamimura and collaborators showed that one can cause the contraction of the legs of a mouse by selectively stimulating specific brain regions by sound \cite{Kamimura2016}. However, it is widely un-understood why ultrasound actually can stimulate nerves. It seems obvious that nerves must be susceptible to mechanical perturbations. Several authors have proposed that the soliton theory could provide a way for explaining the phenomenon (e.g., \cite{Tufail2011, Mueller2014, Feng2024}).
Typical frequencies of ultrasound are 500kHz-5 MHz, corresponding to a wavelength 0.3 to 3 mm in water. This is of a similar length than the width of an action potential in non-myelinated neurons (table \ref{tab:prop_nerv_1}). Ye et al. \cite{Ye2016} showed that lower frequencies are more efficient in stimulating the brain. At such frequencies the wave lengths are closer to the length of action potentials.

\begin{figure}[htbp]
\centering
\includegraphics[width=169pt,height=154pt]{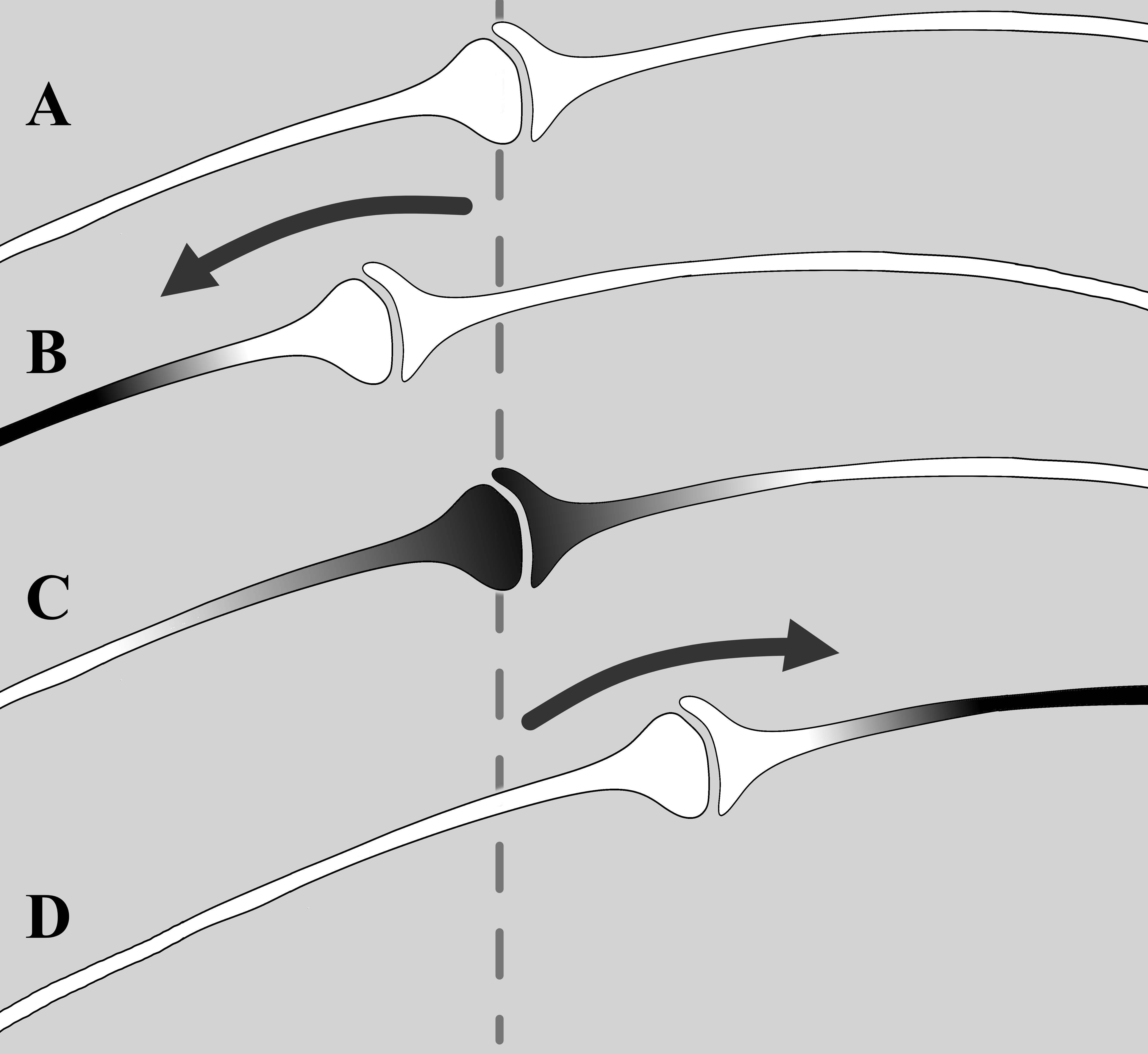}
\caption{The mechanical synapse: Schematic drawing of the motion of a soliton traveling from the left to the right hand side across a synapse and the resulting motion of the synapse. The dark regions represent the location of the soliton (action potential) but its length is largely underrepresented due to graphic limitations. A. In the absence of an excitation. B. A soliton propagates from the left side. Due to the contraction on the presynaptic side, the synapse moves to the left. C. Motion of a soliton across the synapse. The position of the synapse is centered. D. The soliton continues propagating to the right hand side. Due to contraction on the postsynaptic side, the synapse moves towards the right. The overall motion is expected to be much larger than the width of the synaptic gap.}
\label{synapse_motion}
\end{figure}

Contraction of nerves may play a role in signal transmission in the brain. The synaptic gap of a chemical synapse is about 20--40 nm \cite{Kandel2000} and that of an electrical synapse around 2--4 nm wide \cite{Hormuzdi2004}. Let us assume that this distance does not change very much during the transmission of the pulse across the gap. The nerve pulse of non-myelinated fibers has a length of a few mm. Thus, the scale of the nerve pulse can be six or more orders of magnitude larger than the width of the gap (see table \ref{tab:prop_nerv_1}). It can even be larger than the size of small neurons as demonstrated in section \ref{velocityandlengthofthenervepulse}. We have shown above that exited nerves contract when they are not fixed at the ends. Depending on the overall length of the nerve pulse, this can lead to a contraction of hundreds of micrometers or even millimeters in the soliton theory. Experimental evidence shows contraction on a smaller scale but still larger than the width of the synaptic gap. This implies that synapses will move on scales (much) larger than the width of the synaptic gap. In a hydrodynamic picture of the nerve pulse, the pulse might not see a synapse at all and will cross the distance of the synaptic gap with the speed of the soliton because the synapse itself is part of a macroscopic motion. Fig. \ref{synapse_motion} shows the position of a single synapse in the absence and the presence of a moving pulse (the compressed region is shown in dark shades). On expects a motion first towards the presynaptic side followed by motion towards the postsynaptic side. For a non-myelinated nerve with a pulse velocity of 1 m\slash s, the distance of the chemical synaptic gap would be crossed in 20--40 ns, i.e., basically without delay.

The chemical synapse discussed in neuroscience is governed by several sequential processes such as calcium-channel channel open-time in the ms range, membrane fusion processes that should be of the order of the relaxation time of membranes (a few ms at physiological temperature) and by diffusion processes. Thus, the synaptic transmission is expected to be relatively slow - slower than what is observed in many transmission experiments. Greengard \cite{Greengard2001} reports transmission times below 1 ms \cite{Greengard2001}, i.e., faster than the typical open life-time of an ion channel. The time for a synaptic fusion event is estimated to be larger than 0.5 ms judging from relaxation time scales of membranes \cite{Grabitz2002, Seeger2007}. However, synaptic transmission until the postsynaptic signal is about 200 \textmu s in giant squid axons \cite{Holz1999}. Synaptic calyx junctions that connect hair cells to the brain transmit signals faster than any other in the human body \cite{McCue1994a, McCue1994b}. Transmission times below 150 \textmu s have been reported \cite{Borst1996}, faster than it can reasonably be understood by a chemical synapse. Therefore, a non-quantal transmission scheme for such processes was proposed \cite{Govindaraju2023}. Electrical synapses are typically much faster than chemical synapse. The mechanical transmission process proposed here could be considered a `fast mechanical synapse'.
It would be of high interest to test experimentally whether motion of the above kind (hundreds of micrometers to millimeters) can be observed in an active brain. It should be possible to measure this with high resolution imaging methods. One possibility is magnetic resonance elastography (MRE), a new method that allows to measure the stiffness of brain regions, and small motion \cite{Sack2023}. Remember that a nerve soliton consists of a solid membrane region traveling in a fluid environment. An recent application of this technique shows that active brain regions are in fact stiffer, i.e., that the mechanical properties of these regions are altered \cite{Palnitkar2024}. There may exist different possibilities to explain this phenomenon, but it is consistent with the soliton picture.

\begin{figure}[htbp]
\centering
\includegraphics[width=225pt,height=158pt]{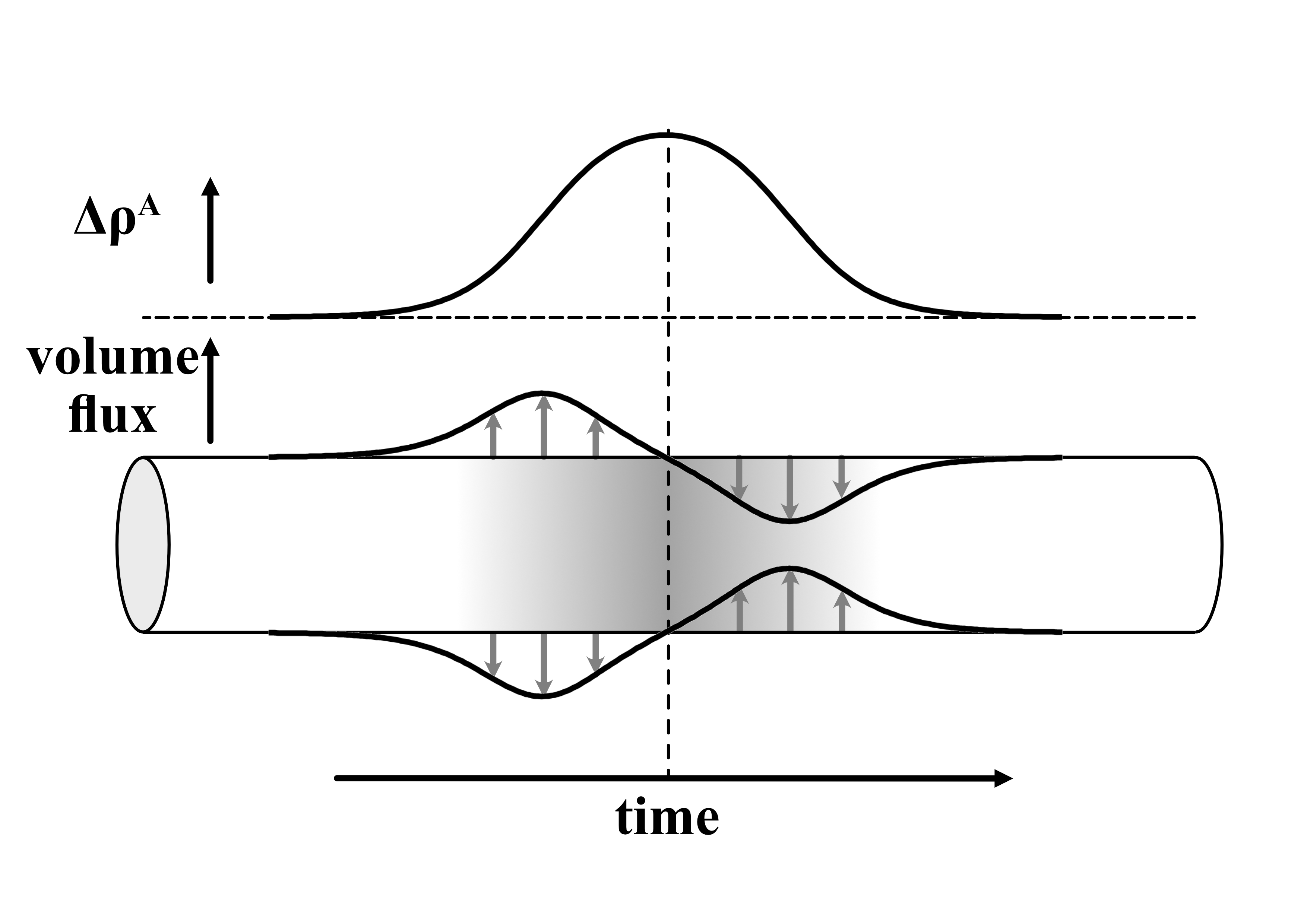}
\caption{The membrane in the transition regime (in the excited region) is permeable \cite{Blicher2009}. If the radius of the nerve axon stays about constant and the length of the axon contracts, one expects electrolyte flowing first out and then back into axon. The membrane outside the soliton has a low permeability and no such flux is expected. }
\label{water_flow_diff_c}
\end{figure}

In section \ref{nervecontractioninthesolitontheory} we discussed the contraction of nerves und conditions of constant volume and of constant radius. Experiments indicate that radius changes are very small while the change in area can be of the order of 20\%. Thus, the volume of the axon changes, resulting in flows of electrolyte out and back into the neuron. According to eq. \ref{eq:theor_contraction04} the volume of a cylindrical element of an axon is given by $V_0=\pi r_0^2 \cdot x_0$ and the surface area is $A_0=2\pi r_0\cdot x_0$. Further, $d x/x_0=-d\rho^A/\rho^A$ at constant radius. When the axon contracts, the volume flux density $J_V$ is therefore given by
\begin{equation}\label{eq:flux01}
	J_V=\frac{1}{A_0} \frac{dV}{dt}=\frac{ r_0}{2  x_0}\frac{dx}{dt}\approx -\frac{r_0}{2\rho_0^A}\frac{d\rho^A}{dt}
\end{equation}
with units {[m\slash s]}. The volume flux density can be deduced from the derivative of the curve in Fig. \ref{soliton_v0655} where the density is given as a function of $z=x-vt$ of a coordinate system moving with the velocity of a soliton. It follows that $d\rho^A/dt=-v\cdot d\rho^A/dz$ and
\begin{equation}\label{eq:flux01}
	J_V= \frac{r_0 v}{2\rho_0^A}\frac{d\rho^A}{dz} \;.
\end{equation}
Thus the flux of electrolyte out and back into the axon is proportional to the radius of the axon and the velocity of the soliton. The flux as a function of time is schematically shown in Fig. \ref{water_flow_diff_c}. As a consequence of contraction one may expect electrical transmembrane currents and associated changes in transmembrane voltage.

\section{Conclusion}
\label{conclusion}

We reviewed some of the evidence for mechanical changes in nerves and neurons. It has been shown experimentally that nerves both contract and change their thickness under the influence of the action potential, and the consequences for neural function. These findings are consistent with the predictions of the soliton theory that considers the nerve pulse as a region of nerve membrane in a more ordered phase state traveling in a fluid membrane. We demonstrate that the nerve pulse is very large, i.e., on the order of millimeters to many centimeters. It is larger than many neurons and orders of magnitude larger than the synaptic gap. Mechanical changes are therefore not of anecdotal nature but may play an important role in the function of the brain, e.g., in synaptic transmission. Contraction of excited nerves is larger than the width of the synaptic gap suggesting that synapses may move during transmission of a pulse. Since the motion of the synapses is just part of a macroscopic motion, nerve pulses might cross synapses without delay. This gives rise to the possibility of \emph{fast mechanical synapses}. It suggests that nerve networks as a whole transport sound-like phenomena. This may explain why it is possible to stimulate nerve activity and influence brain activity by ultra-sound which has typical wave lengths of 0.3--3 mm, which is just the range of the length of the action potentials in living nerves. We further show that contraction may give rise to reversible flows of electrolyte across the membrane caused by changes in the surface to volume ratio of the axon under the influence of the action potential.

\vspace{1cm} \noindent
\textbf{Author contributions:}
TH designed and wrote the article.
\vspace{0.5cm}

\small{
}

\end{document}